\documentclass[a4paper,11pt]{article}

\usepackage{hyperref}%
\hypersetup{%
  colorlinks=true,%
  linkcolor=blue,%
  citecolor=magenta,%
  urlcolor=black,%
  bookmarksnumbered=true,%
  bookmarkstype=toc,%
  bookmarksopen=false,%
  pdftitle={Quantum Walk on Orbit Spaces},%
  pdfkeywords={continuous-time quantum walk; discrete-time quantum walk; orbit space; unitary representation; boundary condition; identical particles},%
  pdfauthor={Satoshi Ohya}}%

\usepackage[margin=1in]{geometry}%

\usepackage[tt=false,sb]{libertine}%
\usepackage[T1]{fontenc}%
\usepackage{libertinust1math}%
\usepackage[cal=stix,scr=boondoxo,bb=boondox]{mathalpha}%
\usepackage{bm}%
\usepackage{mathtools}%
\DeclareMathOperator{\e}{e}%
\DeclareMathOperator{\im}{Im}%
\DeclareMathOperator{\re}{Re}%
\DeclareMathOperator{\sgn}{sgn}%
\DeclareMathOperator{\tr}{tr}%
\allowdisplaybreaks[1]%

\usepackage{epic,eepic,xcolor,graphicx}%

\usepackage[font=small,hang]{caption}%
\usepackage[labelformat=simple]{subcaption}%
\usepackage[backend=biber,style=phys,biblabel=brackets,block=ragged,eprint=true,url=true]{biblatex}%
\title{\Large\bfseries Quantum Walk on Orbit Spaces}%
\author{\normalsize Satoshi Ohya\\[1em]
  \small\itshape Institute of Quantum Science, Nihon University,\\
  \small\itshape Kanda-Surugadai 1-8-14, Chiyoda, Tokyo 101-8308, Japan\\[1ex]
  \small\ttfamily ohya.satoshi@nihon-u.ac.jp}%
\date{\small(Dated: \today)}%

\begin{document}
\maketitle%
\flushbottom%

\begin{abstract}
  Inspired by the covering-space method in path integral on multiply
  connected spaces, we here present a universal formula of
  time-evolution kernels for continuous- and discrete-time quantum
  walks on orbit spaces. In this note, we focus on the case in which
  walkers' configuration space is the orbit space $\Lambda/\Gamma$,
  where $\Lambda$ is an arbitrary lattice and $\Gamma$ is a discrete
  group whose action on $\Lambda$ has no fixed points. We show that
  the time-evolution kernel on $\Lambda/\Gamma$ can be written as a
  weighted sum of time-evolution kernels on $\Lambda$, where the
  summation is over the orbit of initial point in $\Lambda$ and weight
  factors are given by a one-dimensional unitary representation of
  $\Gamma$. Focusing on one dimension, we present a number of examples
  of the formula. We also present universal formulas of resolvent
  kernels, canonical density matrices, and unitary representations of
  arbitrary groups in quantum walks on $\Lambda/\Gamma$, all of which
  are constructed in exactly the same way as for the time-evolution
  kernel.
\end{abstract}

\begingroup%
\hypersetup{linkcolor=black}%
\tableofcontents
\endgroup%

\newpage
\section{Introduction}
\label{section:1}
Quantum walk---a quantum-mechanical analog of classical random walk on
lattices or graphs---has been the subject of intense study over the
last two decades. Just as in classical random walk, there exist two
distinct formulations in quantum walk: continuous-time quantum walk
and discrete-time quantum walk, the former is equivalent to
tight-binding models in condensed matter physics, while the latter is
a natural generalization of classical random walk and formulated
without recourse to Hamiltonian operators. These two formulations have
their own merits and their applications now appear in many
disciplines, including quantum search algorithm
\cite{Shenvi:2002,Childs:2003}, universal quantum computation
\cite{Childs:2008,Lovett:2009,Underwood:2010,Childs:2012}, and
topological phases of matter \cite{Kitagawa:2010}; see
refs.~\cite{Kempe:2003,Venegas-Andraca:2012zkr,Wu:2019} for
reviews. In both formulations, the central object is the probability
amplitude for finding particles (walkers), which is given by a matrix
element of time-evolution operator in position space---the
time-evolution kernel.\footnote{The term ``kernel'' is a remnant of
  continuum theory. In quantum mechanics on continuous spaces, a
  matrix element of time-evolution operator is given by an integral
  kernel.} This time-evolution kernel is normally calculated through
spectral decomposition or numerical calculation, which becomes harder
as the matrix size becomes larger. It would therefore be desirable if
a simpler method existed.

The purpose of this note is to present such a method by generalizing
the Dowker's covering-space method \cite{Dowker:1972np} in path
integral (see also
refs.~\cite{Schulman:1968yv,Laidlaw:1970ei,Horvathy:1979qk,Berg:1981ix,Horvathy:1988vh}). As
is well known, in quantum mechanics on continuous spaces, the
time-evolution kernel can be represented by the Feynman path integral,
which provides a number of powerful methods to analyze quantum systems
nonperturbatively. Among them is the covering-space method: it
provides a universal method to construct the time-evolution kernel on
multiply connected spaces of the form
$\mathcal{M}=\widetilde{\mathcal{M}}/\pi_{1}(\mathcal{M})$, where
$\widetilde{\mathcal{M}}$ is the universal covering space of
$\mathcal{M}$ and $\pi_{1}(\mathcal{M})$ is the fundamental group of
$\mathcal{M}$. In this method, the path integral on $\mathcal{M}$ is
given by a linear combination of partial amplitudes, where each
partial amplitude is given by the path integral on the universal
covering space $\widetilde{\mathcal{M}}$ and linear-combination
coefficients are given by a one-dimensional unitary representations of
the fundamental group $\pi_{1}(\mathcal{M})$. Inspired by this method,
we here present a universal formula for the time-evolution kernel in
both continuous- and discrete-time quantum walks where walkers'
configuration space can be regarded as the \textit{orbit space}
$\Lambda/\Gamma$. Here $\Lambda$ is an arbitrary lattice and $\Gamma$
is a discrete group whose action on $\Lambda$ has no fixed points. A
typical example for such configuration spaces is that for a single
walker on a periodic lattice. Another typical example is the
configuration space for identical walkers on an arbitrary lattice,
where the indistinguishability of identical particles makes their
configuration space an orbit space
\cite{Souriau:1967,Souriau:1969,Laidlaw:1970ei,Leinaas:1977fm,Harshman:2021jlv}. We
show that the time-evolution kernel on the orbit space
$\Lambda/\Gamma$ can be written as a weighted sum of time-evolution
kernels on $\Lambda$, where the summation is over the orbit of initial
point in $\Lambda$ and weight factors are given by a one-dimensional
unitary representation of $\Gamma$. This universal formula offers a
simpler method to construct the time-evolution kernel on
$\Lambda/\Gamma$ because computation becomes generally much easier on
$\Lambda$.

In what follows, we first set up the problem and then present our main
formula and its proof. We then present a number of examples of the
formula in section~\ref{section:3}. In section~\ref{section:4}, we
present several other quantities that can be constructed in exactly
the same way as for the time-evolution kernel. Examples include the
resolvent kernel, the canonical density matrix, and a unitary
representation of arbitrary groups. Section~\ref{section:5} is devoted
to the conclusion. Appendix \ref{appendix:A} presents some sample
computations in continuous-time quantum walk.

Throughout this note we will use the units in which $\hbar=a=1$, where
$a$ is a lattice spacing.

\section{Time-evolution kernel}
\label{section:2}
To begin with, let us fix some notation. Let $\Lambda$ be an arbitrary
lattice (i.e., a discrete space spanned by a set of linearly
independent vectors in a Euclidean space) and let $\Gamma$ be a
discrete group whose action on $\Lambda$ has no fixed points. We note
that $\Gamma$ must be a discrete subgroup of the isometry of the
Euclidean space, which consists of reflections, translations, and
rotations. Let $\Lambda/\Gamma$ be the orbit space (quotient space)
given by the identification $x\sim\gamma x$ in $\Lambda$, where
$\gamma x$ stands for the action of $\gamma\in\Gamma$ on $x\in\Lambda$
that satisfies the compatibility condition
$\gamma_{1}(\gamma_{2}x)=(\gamma_{1}\gamma_{2})x$ for any
$\gamma_{1},\gamma_{2}\in\Gamma$ and $x\in\Lambda$. For the moment, we
shall consider continuous-time quantum walk on the lattice
$\Lambda/\Gamma$, where the Hilbert space $\mathscr{H}$ is the set of
square-summable sequences on $\Lambda/\Gamma$,
$\mathscr{H}=l^{2}(\Lambda/\Gamma)$. (Note, however, that the formula
presented below is turned out to be applicable to discrete-time
quantum walk as well; see section~\ref{section:4.3}.) The action of
the time-evolution operator $U_{\tau}$ on a state
$\psi_{0}\in\mathscr{H}$ is defined by
\begin{align}
  (U_{\tau}\psi_{0})(x)\coloneq\sum_{y\in\Lambda/\Gamma}U_{\tau}(x,y)\psi_{0}(y),\quad\forall x\in\Lambda/\Gamma,\label{eq:1}
\end{align}
where $U_{\tau}(x,y)$ is the time-evolution kernel and the subscript
$\tau\in\mathbb{R}$ represents the time. The probability for finding a
particle at the time $\tau$ and at the position $x$ is then given by
\begin{align}
  P_{\tau}(x)=|(U_{\tau}\psi_{0})(x)|^{2}.\label{eq:2}
\end{align}
In particular, if the particle is initially localized at $x=x_{0}$
(i.e., $\psi_{0}(x)=\delta_{x,x_{0}}$), the probability is simply
given by $P_{\tau}(x)=|U_{\tau}(x,x_{0})|^{2}$.

In the following, we shall construct $U_{\tau}(x,y)$ in terms of the
time-evolution kernel on $\Lambda$. The key is the group property of
the time-evolution operator.

\subsection{The formula}
\label{section:2.1}
The time-evolution operator $U_{\tau}$ is a one-parameter family of
unitary operators. It satisfies the composition law
$U_{\tau_{1}}U_{\tau_{2}}=U_{\tau_{1}+\tau_{2}}$, the unitarity
$U_{\tau}^{\dagger}(=U_{\tau}^{-1})=U_{-\tau}$, and the initial
condition $U_{0}=I$, where $I$ stands for the identity operator.
Correspondingly, the time-evolution kernel $U_{\tau}(\cdot,\cdot)$
must satisfy the following properties:
\begin{subequations}
  \begin{itemize}
  \item\textbf{Property 1.~(Composition law)}
    \begin{align}
      \sum_{z\in\Lambda/\Gamma}U_{\tau_{1}}(x,z)U_{\tau_{2}}(z,y)=U_{\tau_{1}+\tau_{2}}(x,y),\quad\forall x,y\in\Lambda/\Gamma.\label{eq:3a}
    \end{align}
  \item\textbf{Property 2.~(Unitarity)}
    \begin{align}
      \overline{U_{\tau}(x,y)}=U_{-\tau}(y,x),\quad\forall x,y\in\Lambda/\Gamma.\label{eq:3b}
    \end{align}
  \item\textbf{Property 3.~(Initial condition)}
    \begin{align}
      U_{0}(x,y)=\delta_{x,y},\quad\forall x,y\in\Lambda/\Gamma.\label{eq:3c}
    \end{align}
  \end{itemize}
\end{subequations}
Here the overline ($\overline{\phantom{m}}$) stands for the complex
conjugate. As we shall prove shortly, such a kernel can be constructed
as follows:
\begin{align}
  U_{\tau}(x,y)=\sum_{\gamma\in\Gamma}D(\gamma)\widetilde{U}_{\tau}(x,\gamma y),\label{eq:4}
\end{align}
where $D:\Gamma\to U(1)$ ($\gamma\mapsto D(\gamma)$) is a
one-dimensional unitary representation of $\Gamma$ that satisfies the
group composition law
$D(\gamma)D(\gamma^{\prime})=D(\gamma\gamma^{\prime})$ and the
unitarity $\overline{D(\gamma)}=D(\gamma)^{-1}=D(\gamma^{-1})$ for any
$\gamma,\gamma^{\prime}\in\Gamma$.  Here
$\widetilde{U}_{\tau}(\cdot,\cdot)$ is a time-evolution kernel on
$\Lambda$ that satisfies the following assumptions:
\begin{subequations}
  \begin{itemize}
  \item\textbf{Assumption 1.~(Composition law)}
    \begin{align}
      \sum_{z\in\Lambda}\widetilde{U}_{\tau_{1}}(x,z)\widetilde{U}_{\tau_{2}}(z,y)=\widetilde{U}_{\tau_{1}+\tau_{2}}(x,y),\quad\forall x,y\in\Lambda.\label{eq:5a}
    \end{align}
  \item\textbf{Assumption 2.~(Unitarity)}
    \begin{align}
      \overline{\widetilde{U}_{\tau}(x,y)}=\widetilde{U}_{-\tau}(y,x),\quad\forall x,y\in\Lambda.\label{eq:5b}
    \end{align}
  \item\textbf{Assumption 3.~(Initial condition)}
    \begin{align}
      \widetilde{U}_{0}(x,y)=\delta_{x,y},\quad\forall x,y\in\Lambda.\label{eq:5c}
    \end{align}
  \item\textbf{Assumption 4.~($\Gamma$-invariance)}
    \begin{align}
      \widetilde{U}_{\tau}(\gamma x,\gamma y)=\widetilde{U}_{\tau}(x,y),\quad\forall x,y\in\Lambda,\quad\forall\gamma\in\Gamma.\label{eq:5d}
    \end{align}
  \end{itemize}
\end{subequations}
We note that the $\Gamma$-invariance \eqref{eq:5d} is guaranteed if
the Hamiltonian operator on $\Lambda$ is invariant under the action of
$\Gamma$.

Before giving the proof, let us first present a quick derivation of
formula \eqref{eq:4} by following the Dowker method
\cite{Dowker:1972np}. To this end, let $\widetilde{\psi}_{\tau}(x)$ be
an equivariant function on $\Lambda$ that satisfies
$\widetilde{\psi}_{\tau}(\gamma
x)=D(\gamma)\widetilde{\psi}_{\tau}(x)$ for any $x\in\Lambda$ and
$\gamma\in\Gamma$. (The reason for using this will be apparent
shortly.)  Then we have
\begin{align}
  \widetilde{\psi}_{\tau}(x)
  &=\sum_{y\in\Lambda}\widetilde{U}_{\tau}(x,y)\widetilde{\psi}_{0}(y)\nonumber\\
  &=\sum_{y\in\Lambda/\Gamma}\sum_{\gamma\in\Gamma}\widetilde{U}_{\tau}(x,\gamma y)\widetilde{\psi}_{0}(\gamma y)\nonumber\\
  &=\sum_{y\in\Lambda/\Gamma}\sum_{\gamma\in\Gamma}\widetilde{U}_{\tau}(x,\gamma y)D(\gamma)\widetilde{\psi}_{0}(y)\nonumber\\
  &=\sum_{y\in\Lambda/\Gamma}\left(\sum_{\gamma\in\Gamma}D(\gamma)\widetilde{U}_{\tau}(x,\gamma y)\right)\widetilde{\psi}_{0}(y),\label{eq:6}
\end{align}
where in the second equality we have used the following identity:
\begin{align}
  \sum_{x\in\Lambda}f(x)=\sum_{x\in\Lambda/\Gamma}\sum_{\gamma\in\Gamma}f(\gamma x).\label{eq:7}
\end{align}
Here $f(x)$ is an arbitrary test function on $\Lambda$. This identity
just says that first summing over the orbit
$\Gamma\cdot x\coloneq\{\gamma x:\gamma\in\Gamma\}$ of
$x\in\Lambda/\Gamma$ and then summing over all $x\in\Lambda/\Gamma$
yields the summation over the whole space $\Lambda$.  By comparing
eq.~\eqref{eq:6} with definition \eqref{eq:1}, we arrive at formula
\eqref{eq:4}.

Now, since $\widetilde{U}_{\tau}(\cdot,\cdot)$ is defined on the
lattice $\Lambda$, the domain of $U_{\tau}(\cdot,\cdot)$ defined by
eq.~\eqref{eq:4} can be naturally extended from $\Lambda/\Gamma$ to
$\Lambda$. In particular, it satisfies the following equation:
\begin{align}
  U_{\tau}(\gamma x,y)=D(\gamma)U_{\tau}(x,y),\quad\forall x,y\in\Lambda,\quad\forall\gamma\in\Gamma.\label{eq:8}
\end{align}
In fact, a straightforward calculation gives
\begin{align}
  U_{\tau}(\gamma x,y)
  &=\sum_{\gamma^{\prime}\in\Gamma}D(\gamma^{\prime})\widetilde{U}_{\tau}(\gamma x,\gamma^{\prime}y)\nonumber\\
  &=\sum_{\gamma^{\prime}\in\Gamma}D(\gamma\gamma^{-1}\gamma^{\prime})\widetilde{U}_{\tau}(\gamma^{-1}\gamma x,\gamma^{-1}\gamma^{\prime}y)\nonumber\\
  &=D(\gamma)\sum_{\gamma^{\prime}\in\Gamma}D(\gamma^{-1}\gamma^{\prime})\widetilde{U}_{\tau}(x,\gamma^{-1}\gamma^{\prime}y)\nonumber\\
  &=D(\gamma)\sum_{\gamma^{\prime\prime}\in\Gamma}D(\gamma^{\prime\prime})\widetilde{U}_{\tau}(x,\gamma^{\prime\prime}y)\nonumber\\
  &=D(\gamma)U_{\tau}(x,y),\label{eq:9}
\end{align}
where the second equality follows from the $\Gamma$-invariance
\eqref{eq:5d} and the third equality follows from the group
composition law
$D(\gamma\gamma^{-1}\gamma^{\prime})=D(\gamma)D(\gamma^{-1}\gamma^{\prime})$. In
the fourth equality, we have changed the summation variable from
$\gamma^{\prime}$ to
$\gamma^{\prime\prime}\coloneq\gamma^{-1}\gamma^{\prime}$. It is now
obvious from eq.~\eqref{eq:8} that $(U_{\tau}\psi_{0})(x)$ defined by
eq.~\eqref{eq:1} also satisfies
$(U_{\tau}\psi_{0})(\gamma x)=D(\gamma)(U_{\tau}\psi_{0})(x)$ for any
$x\in\Lambda$ and $\gamma\in\Gamma$; that is, $(U_{\tau}\psi_{0})(x)$
becomes an equivariant function on $\Lambda$. This is the reason why
we used the equivariant function in the above derivation. As we shall
see in section~\ref{section:3}, eq.~\eqref{eq:8} provides boundary
conditions on $\Lambda/\Gamma$.

Finally, let us comment on the case where the action of $\Gamma$ has
fixed points. First, identity \eqref{eq:7} does not hold in general if
there is a fixed point: if there is a point $x\in\Lambda$ that
satisfies $\gamma x=x$ for some $\gamma(\neq e)\in\Gamma$, where $e$
stands for the identity element of $\Gamma$, the right-hand side of
eq.~\eqref{eq:7} leads to an overcounting of the fixed point
$x$.\footnote{In general, eq.~\eqref{eq:7} becomes
  $\sum_{x\in\Lambda}f(x)=\sum_{x\in(\Lambda-\Delta)/\Gamma}\sum_{\gamma\in\Gamma}f(x)+\sum_{x\in\Delta}f(x)$,
  where $\Delta$ stands for the set of fixed points of $\Gamma$.}
Note, however, that if $f(x)$ is subject to the Dirichlet boundary
condition at the fixed point, such an overcounting does not occur so
that eq.~\eqref{eq:7} holds true even in the presence of fixed
points.\footnote{More generally, such an overcounting does not occur
  if $\sum_{x\in\Delta}f(x)=0$.} Note that the Dirichlet boundary
condition $f(x)=0$ at $x=\gamma x$ can be deduced from the equivariant
property $D(\gamma)f(x)=f(\gamma x)=f(x)$ if $D(\gamma)\neq1$. Hence,
if $D:\Gamma\to U(1)$ is not the trivial representation, our formula
\eqref{eq:4} can be applied equally well to the case in which the
action of $\Gamma$ has fixed points. For the case of the trivial
representation, however, the equivariant property does not lead to any
definite boundary conditions. For simplicity, in this note we will
mainly focus on the case where $\Gamma$ has no fixed points.

\subsection{Proof}
\label{section:2.2}
Now we show that $U_{\tau}(\cdot,\cdot)$ given by formula \eqref{eq:4}
satisfies the required properties \eqref{eq:3a}--\eqref{eq:3c} if $D$
is a one-dimensional unitary representation of $\Gamma$ and if
$\widetilde{U}_{\tau}(\cdot,\cdot)$ satisfies the assumptions
\eqref{eq:5a}--\eqref{eq:5d}. The proof is by direct computation. Each
property is proved as follows. (See also
refs.~\cite{Ohya:2011qu,Ohya:2012qj,Ohya:2021oni} for similar proofs
in path integral.)

\paragraph*{Property 1. (Composition law)}
Let us first prove the composition law \eqref{eq:3a}. By substituting
eq.~\eqref{eq:4} into the left-hand side of eq.~\eqref{eq:3a}, we get
\begin{align}
  \sum_{z\in\Lambda/\Gamma}U_{\tau_{1}}(x,z)U_{\tau_{2}}(z,y)
  &=\sum_{z\in\Lambda/\Gamma}\sum_{\gamma_{1}\in\Gamma}\sum_{\gamma_{2}\in\Gamma}D(\gamma_{1})D(\gamma_{2})\widetilde{U}_{\tau_{1}}(x,\gamma_{1}z)\widetilde{U}_{\tau_{2}}(z,\gamma_{2}y)\nonumber\\
  &=\sum_{z\in\Lambda/\Gamma}\sum_{\gamma_{1}\in\Gamma}\sum_{\gamma_{2}\in\Gamma}D(\gamma_{1}\gamma_{2})\widetilde{U}_{\tau_{1}}(x,\gamma_{1}z)\widetilde{U}_{\tau_{2}}(\gamma_{1}z,\gamma_{1}\gamma_{2}y)\nonumber\\
  &=\sum_{\gamma\in\Gamma}D(\gamma)\sum_{z\in\Lambda/\Gamma}\sum_{\gamma_{1}\in\Gamma}\widetilde{U}_{\tau_{1}}(x,\gamma_{1}z)\widetilde{U}_{\tau_{2}}(\gamma_{1}z,\gamma y)\nonumber\\
  &=\sum_{\gamma\in\Gamma}D(\gamma)\sum_{z\in\Lambda}\widetilde{U}_{\tau_{1}}(x,z)\widetilde{U}_{\tau_{2}}(z,\gamma y)\nonumber\\
  &=\sum_{\gamma\in\Gamma}D(\gamma)\widetilde{U}_{\tau_{1}+\tau_{2}}(x,\gamma y)\nonumber\\
  &=U_{\tau_{1}+\tau_{2}}(x,y),\label{eq:10}
\end{align}
where the second equality follows from the group composition law
$D(\gamma_{1})D(\gamma_{2})=D(\gamma_{1}\gamma_{2})$ and the
$\Gamma$-invariance \eqref{eq:5d}. The third equality follows from the
change of the summation variable from $\gamma_{2}$ to
$\gamma\coloneq\gamma_{1}\gamma_{2}$, and the fourth equality follows
from formula \eqref{eq:7}. Finally, the fifth equality follows from
assumption \eqref{eq:5a}.

\paragraph*{Property 2. (Unitarity)}
Let us next prove unitarity \eqref{eq:3b}. By substituting
eq.~\eqref{eq:4} into the left-hand side of eq.~\eqref{eq:3b}, we get
\begin{align}
  \overline{U_{\tau}(x,y)}
  &=\sum_{\gamma\in\Gamma}\overline{D(\gamma)}\,\overline{\widetilde{U}_{\tau}(x,\gamma y)}\nonumber\\
  &=\sum_{\gamma\in\Gamma}D(\gamma^{-1})\widetilde{U}_{-\tau}(\gamma y,x)\nonumber\\
  &=\sum_{\gamma\in\Gamma}D(\gamma^{-1})\widetilde{U}_{-\tau}(y,\gamma^{-1}x)\nonumber\\
  &=U_{-\tau}(y,x),\label{eq:11}
\end{align}
where the second equality follows from the unitarity properties
$\overline{D(\gamma)}=D(\gamma^{-1})$ and \eqref{eq:5b}. The third
equality follows from the $\Gamma$-invariance \eqref{eq:5d}, and the
last equality follows from definition \eqref{eq:4} (where the
summation is over $\gamma^{-1}$ instead of $\gamma$).

\paragraph*{Property 3. (Initial condition)}
Let us finally prove the initial condition \eqref{eq:3c}. By
substituting eq.~\eqref{eq:4} into the left-hand side of
eq.~\eqref{eq:3c}, we get
\begin{align}
  U_{0}(x,y)
  &=\sum_{\gamma\in\Gamma}D(\gamma)\widetilde{U}_{0}(x,\gamma y)\nonumber\\
  &=\sum_{\gamma\in\Gamma}D(\gamma)\delta_{x,\gamma y}\nonumber\\
  &=D(e)\delta_{x,e y}\nonumber\\
  &=\delta_{x,y},\label{eq:12}
\end{align}
where the second equality follows from assumption \eqref{eq:5c}. The
third equality follows from the fact that $x$ and $\gamma y$ cannot be
equal for any $x,y\in\Lambda/\Gamma$ except for the case
$\gamma=e$. Finally, the last equality follows from $D(e)=1$ for any
one-dimensional unitary representations of $\Gamma$.

\bigskip

Putting all the above things together, we see that eq.~\eqref{eq:4} is
the sufficient condition to be the time-evolution kernel on the orbit
space $\Lambda/\Gamma$. This completes the proof.

\section{Examples}
\label{section:3}
There exist a number of examples in which walkers' configuration space
can be regarded as an orbit space. Typical examples are a single
walker on a torus, the half space, and a cubic. Another typical
example is identical walkers on an arbitrary lattice, where their
configuration space always becomes an orbit space. In this section, we
shall focus on one spatial dimension for simplicity and present
several examples that fit into formula \eqref{eq:4}. Let us start with
single-walker examples.

\subsection{A single walker in one dimension}
\label{section:3.1}
Let $\widetilde{U}_{\tau}(x,y)$ be a time-evolution kernel on the
integer lattice $\Lambda=\mathbb{Z}$ that satisfies the composition
law \eqref{eq:5a}, the unitarity \eqref{eq:5b}, and the initial
condition \eqref{eq:5c} as well as the translation invariance
$\widetilde{U}_{\tau}(x+z,y+z)=\widetilde{U}_{\tau}(x,y)$ and the
reflection invariance
$\widetilde{U}_{\tau}(z-x,z-y)=\widetilde{U}_{\tau}(x,y)$ for any
$x,y,z\in\mathbb{Z}$. A typical example of such a kernel is that of a
free particle given by
$\widetilde{U}_{\tau}(x,y)=\e^{i\frac{\pi}{2}|x-y|}J_{|x-y|}(\omega\tau)$,
where $J_{n}$ is the Bessel function of the first kind and
$\omega(>0)$ is a hopping parameter; see eq.~\eqref{eq:A.8} in
appendix \ref{appendix:A}. (Note, however, that the formulas presented
below are not limited to free-particle theories. They are robust
against any perturbations unless boundary conditions \eqref{eq:8} are
changed.)  Below we shall construct time-evolution kernels for a
single walker on a circle, the half line, and a finite interval by
gauging these discrete symmetries.

\begin{figure}[t!]
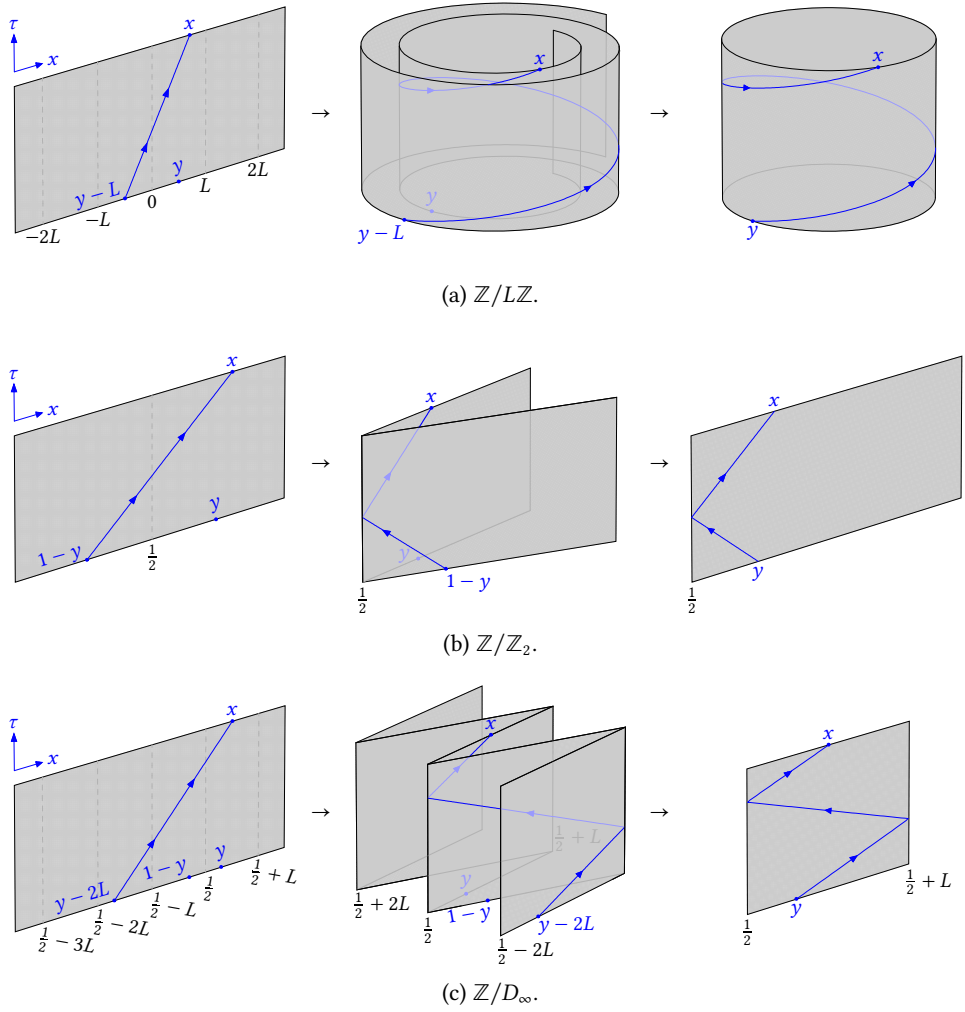

  \centering%
  \begin{subfigure}[t]{\linewidth}
    \centering%
    \input{figure1a.eepic}%
    \caption{$\mathbb{Z}/L\mathbb{Z}$.}
    \label{figure:1a}
  \end{subfigure}
  \begin{subfigure}[t]{\linewidth}
    \centering%
    \input{figure1b.eepic}%
    \caption{$\mathbb{Z}/\mathbb{Z}_{2}$.}
    \label{figure:1b}
  \end{subfigure}
  \begin{subfigure}[t]{\linewidth}
    \centering%
    \input{figure1c.eepic}%
    \caption{$\mathbb{Z}/D_{\infty}$.}
    \label{figure:1c}
  \end{subfigure}
  \caption{Construction of one-dimensional orbit spaces
    $\mathbb{Z}/L\mathbb{Z}$, $\mathbb{Z}/\mathbb{Z}_{2}$, and
    $\mathbb{Z}/D_{\infty}$ and typical single-walker trajectories. In
    the spacetime picture, these orbit spaces correspond to a
    cylinder, the half space, and a finite strip. (a) A cylinder of
    circumference $L$ is constructed by rolling up the infinite
    strip. This is equivalent to making the identification
    $x\sim x+nL$ ($n=0,\pm1,\pm2,\cdots$) in the infinite strip. Under
    this identification, a single-walker trajectory with the initial
    and final points $y+nL$ and $x$ is mapped to a trajectory that
    starts from the initial point $y$ and reaches the final point $x$
    after winding around the cylinder $n$ times in the clockwise
    direction. The walker acquires the Aharonov-Bohm phase
    $\e^{i\theta}$ every time it winds around the cylinder, and this
    phase is described by the unitary representation
    $D^{[\theta]}:L\mathbb{Z}\to U(1)$. (b) The half space with the
    boundary at $x=1/2$ is constructed by folding the infinite strip
    in half at $x=1/2$. This is equivalent to making the
    identification $x\sim1-x$ in the infinite strip. (Note, however,
    that noninteger points are excluded in the integer lattice. Hence,
    the boundary is in fact at $x=1$ in our lattice problem.) Under
    this identification, a single-walker trajectory with the initial
    and final points $1-y$ and $x$ is mapped to a trajectory that
    bounces off the boundary. The walker acquires the phase
    $\e^{i\phi}=\pm1$ every time it hits the boundary, and this phase
    is described by the unitary representation
    $D^{[\phi]}:\mathbb{Z}_{2}\to U(1)$. (c) A finite strip with the
    boundaries at $x=1/2$ and $1/2+L$ is constructed by folding up the
    infinite strip at $x=1/2+nL$ ($n=0,\pm1,\pm2,\cdots$). This is
    equivalent to making the identifications $x\sim x+2nL$ and
    $x\sim 1-x+2nL$ in the infinite strip. Under these
    identifications, a single-particle trajectory with the initial and
    final points $y+2nL$ ($1-y+2nL$) and $x$ is mapped to a trajectory
    that bounces off the boundaries $2n$ ($2n+1$) times. The walker
    acquires the phase $\e^{i\theta}=\pm1$
    ($\e^{i(\theta+\phi)}=\pm1$) every time it hits the left (right)
    boundary, and these phases are described by the unitary
    representation $D^{[\theta,\phi]}:D_{\infty}\to U(1)$.}
  \label{figure:1}
\end{figure}

\paragraph*{Example 1. (A single walker on a circle)}
Let us first consider a single walker on a periodic lattice of $L$
sites, $\{1,2,\cdots,L\pmod L\}$. This lattice can be constructed from
$\mathbb{Z}$ by making the identification $x\sim x+nL$, where $n$ is
an arbitrary integer; see figure~\ref{figure:1a}. Hence the
configuration space is the orbit space $\mathbb{Z}/L\mathbb{Z}$, where
$L\mathbb{Z}=\langle t\mid\emptyset\rangle$ is the free group
generated by a translation $t$. Its action on $\mathbb{Z}$ is defined
by
\begin{align}
  tx\coloneq x+L.\label{eq:13}
\end{align}
Note that any element of $L\mathbb{Z}$ can be written as the product
$t^{n}$, whose action on $\mathbb{Z}$ is given by $t^{n}x=x+nL$.

Now we need to find out one-dimensional unitary representations of
$L\mathbb{Z}$. Since $L\mathbb{Z}$ is the free group generated by a
single generator $t$, we have a one-parameter family of maps
$D^{[\theta]}:L\mathbb{Z}\to U(1)$ labeled by an angle parameter
$\theta$:
\begin{align}
  D^{[\theta]}(t)=\e^{i\theta},\label{eq:14}
\end{align}
where $\theta\in\mathbb{R}/2\pi\mathbb{R}$. It then follows from
formula \eqref{eq:4} that the time-evolution kernel for a single
walker on $\mathbb{Z}/L\mathbb{Z}$ takes the following form:
\begin{align}
  U_{\tau}^{[\theta]}(x,y)
  &=\sum_{n=-\infty}^{\infty}D^{[\theta]}(t^{n})\widetilde{U}_{\tau}(x,t^{n}y)\nonumber\\
  &=\sum_{n=-\infty}^{\infty}\e^{in\theta}\widetilde{U}_{\tau}(x,y+nL).\label{eq:15}
\end{align}
Just as in the path integral on a circle (see, e.g., section 2.4 of
ref.~\cite{Chaichian:2001cz}), eq.~\eqref{eq:15} represents the
summation over winding numbers. Physically, eq.~\eqref{eq:15}
describes the situation in which the walker acquires the Aharonov-Bohm
phase $\e^{i\theta}$ every time it winds around the circle, where
$\theta$ plays the role of a magnetic flux penetrating through the
circle. This is the physical meaning of the weight factor
\eqref{eq:14} and the summation over the orbit of initial point. See
also figure~\ref{figure:1a}.

Now two remarks are in order. First, it follows from eq.~\eqref{eq:8}
that $U_{\tau}^{[\theta]}(\cdot,\cdot)$ satisfies the identity
$U_{\tau}^{[\theta]}(x+L,y)=\e^{i\theta}U_{\tau}^{[\theta]}(x,y)$;
that is, it satisfies the twisted boundary conditions
$U_{\tau}^{[\theta]}(L+1,y)=\e^{i\theta}U_{\tau}^{[\theta]}(1,y)$ and
$U_{\tau}^{[\theta]}(0,y)=\e^{-i\theta}U_{\tau}^{[\theta]}(L,y)$. Namely,
eq.~\eqref{eq:15} gives the universal formula of the time-evolution
kernel for a single walker on a circle subject to these twisted
boundary conditions.

The second remark is that, under the reflection, eq.~\eqref{eq:15}
satisfies
$U_{\tau}^{[\theta]}(z-x,z-y)=U_{\tau}^{[-\theta]}(x,y)$. Hence, at
$\theta=0$ or $\pi\pmod{2\pi}$, eq.~\eqref{eq:15} becomes reflection
invariant. We can use this invariance for the construction of
time-evolution kernels on a finite interval; see example 3.

\paragraph*{Example 2. (A single walker on the half line)}
Let us next consider a single walker on a semi-infinite lattice
$\{1,2,\cdots\}$. This lattice can be constructed from the integer
lattice $\mathbb{Z}$ by making the identification $x\sim1-x$; see
figure~\ref{figure:1b}. Hence the configuration space is the orbit
space $\mathbb{Z}/\mathbb{Z}_{2}$, where
$\mathbb{Z}_{2}=\langle r\mid r^{2}=e\rangle$ is the cyclic group of
order 2. Here $r$ is the reflection whose action on $\mathbb{Z}$ is
defined by
\begin{align}
  rx\coloneq 1-x.\label{eq:16}
\end{align}
Note that $r^{2}x=x$. Note also that reflection \eqref{eq:16} does not
have a fixed point in the integer lattice. (Its fixed point is
$x=1/2$.)

Now, since $r^{2}=e$, any one-dimensional unitary representation
$D:\mathbb{Z}_{2}\to U(1)$ must satisfy the condition $D(r)^{2}=1$,
whose solution is $D(r)=\pm1$. Hence there exist two distinct maps
$D^{[\phi]}$ given by
\begin{align}
  D^{[\phi]}(r)=\e^{i\phi},\label{eq:17}
\end{align}
where $\phi\in\{0,\pi\pmod{2\pi}\}$. Correspondingly, there exist the
following two distinct time-evolution kernels for a single walker on
$\mathbb{Z}/\mathbb{Z}_{2}$:
\begin{align}
  U_{\tau}^{[\phi]}(x,y)
  &=\sum_{n=0}^{1}D^{[\phi]}(r^{n})\widetilde{U}_{\tau}(x,r^{n}y)\nonumber\\
  &=\widetilde{U}_{\tau}(x,y)+\e^{i\phi}\widetilde{U}_{\tau}(x,1-y).\label{eq:18}
\end{align}
Again, just as in the path integral on the half line
\cite{Clark:1980xt,Farhi:1989jz,Ohya:2011qu}, eq.~\eqref{eq:18}
represents the summation over bouncing numbers off the boundary: the
$n=0$ term is the contribution from the direct path, while the $n=1$
term is the contribution from the reflected path off the boundary. The
physical meaning of the weight factor \eqref{eq:17} is now clear: it
plays the role of the reflection amplitude off the boundary. In other
words, the walker acquires the phase shift $\phi$ when reflected from
the boundary. See also figure~\ref{figure:1b}.

Notice that eq.~\eqref{eq:18} satisfies the identity
$U_{\tau}^{[\phi]}(1-x,y)=\e^{i\phi}U_{\tau}^{[\phi]}(x,y)$; that is,
it satisfies the boundary condition
$U_{\tau}^{[\phi]}(0,y)=\e^{i\phi}U_{\tau}^{[\phi]}(1,y)$. Hence,
eq.~\eqref{eq:18} gives the universal form of the time-evolution
kernel for a single walker on the half line subject to this boundary
condition. We emphasize that, as noted at the end of
section~\ref{section:2.1}, if one wants a theory subject to the
Dirichlet boundary condition at $x=0$, one should consider the
reflection defined by $rx\coloneq -x$ and choose the representation
$\phi=\pi$. In this case, one arrives at the formula
$U_{\tau}^{[\phi=\pi]}(x,y)=\widetilde{U}_{\tau}(x,y)-\widetilde{U}_{\tau}(x,-y)$
which satisfies $U_{\tau}^{[\phi=\pi]}(0,y)=0$.

\paragraph*{Example 3. (A single walker on a finite interval)}
Let us next consider a single walker on a finite interval of $L$
sites, $\{1,2,\cdots,L\}$. This lattice can be constructed from
$\mathbb{Z}$ by making the identifications $x\sim x+2nL$ and
$x\sim 1-x+2nL$, where $n$ is an arbitrary integer; see
figure~\ref{figure:1c}. Hence, the configuration space is the orbit
space $\mathbb{Z}/D_{\infty}$, where
$D_{\infty}=\mathbb{Z}\rtimes\mathbb{Z}_{2}=\langle t,r\mid r^{2}=e,
rtr=t^{-1}\rangle$ is the infinite dihedral group generated by a
translation $t$ and a reflection $r$.\footnote{The infinite dihedral
  group can also be written as the free product
  $D_{\infty}\cong\mathbb{Z}_{2}\ast\mathbb{Z}_{2}=\langle
  r,r^{\prime}\mid r^{2}=e,r^{\prime2}=e\rangle$, where
  $r^{\prime}(=tr)$ is another reflection defined by
  $r^{\prime}x\coloneq2L+1-x$.} The actions of these operators on
$\mathbb{Z}$ are defined as follows:
\begin{align}
  tx\coloneq x+2L
  \quad\text{and}\quad
  rx\coloneq 1-x.\label{eq:19}
\end{align}
Note that any element of $D_{\infty}$ can be written as $t^{n}r^{m}$,
where $n=0,\pm1,\pm2,\cdots$ and $m=0,1$. The action of this operator
on $\mathbb{Z}$ is given by $t^{n}r^{m}x=x+2nL$ for $m=0$ and
$t^{n}r^{m}x=1-x+2nL$ for $m=1$, respectively. Note also that, in
contrast to the previous examples, $D_{\infty}$ is a non-Abelian
discrete group.

Now, since $r^{2}=e$ and $rtr=t^{-1}$, any one-dimensional unitary
representation $D:D_{\infty}\to U(1)$ must satisfy the conditions
$D(r)^{2}=1$ and $D(r)D(t)D(r)=D(t)^{-1}$, which leads to
$D(t)^{2}=1$. Thus we have $D(t)=\pm1$ and $D(r)=\pm1$; that is, there
exist $2^{2}=4$ distinct maps $D^{[\theta,\phi]}$ given by
\begin{align}
  D^{[\theta,\phi]}(t)=\e^{i\theta}\quad\text{and}\quad D^{[\theta,\phi]}(r)=\e^{i\phi},\label{eq:20}
\end{align}
where $\theta,\phi\in\{0,\pi\pmod{2\pi}\}$. Correspondingly, there
exist the following four distinct time-evolution kernels for a single
walker on $\mathbb{Z}/D_{\infty}$:
\begin{align}
  U_{\tau}^{[\theta,\phi]}(x,y)
  &=\sum_{n=-\infty}^{\infty}\sum_{m=0}^{1}D^{[\theta,\phi]}(t^{n}r^{m})\widetilde{U}_{\tau}(x,t^{n}r^{m}y)\nonumber\\
  &=\sum_{n=-\infty}^{\infty}\left[\e^{in\theta}\widetilde{U}_{\tau}(x,y+2nL)+\e^{in\theta}\e^{i\phi}\widetilde{U}_{\tau}(x,1-y+2nL)\right].\label{eq:21}
\end{align}
Once again, just as in the path integral on a finite interval
\cite{Janke:1979fv,Inomata:1980th,Goodman:1981,Ohya:2012qj},
eq.~\eqref{eq:21} represents the summation over bouncing numbers off
the two boundaries. Physically, $\e^{i\phi}$ and $\e^{i(\theta+\phi)}$
play the roles of the reflection amplitudes off the boundaries $x=1$
and $x=L$, respectively. See also figure~\ref{figure:1c}.

Now, it follows from eq.~\eqref{eq:8} that eq.~\eqref{eq:21} satisfies
the identities
$U_{\tau}^{[\theta,\phi]}(x+2L,y)=\e^{i\theta}U_{\tau}^{[\theta,\phi]}(x,y)$
and
$U_{\tau}^{[\theta,\phi]}(1-x,y)=\e^{i\phi}U_{\tau}^{[\theta,\phi]}(x,y)$,
which implies the boundary conditions
$U_{\tau}^{[\theta,\phi]}(0,y)=\e^{i\phi}U_{\tau}^{[\theta,\phi]}(1,y)$
and
$U_{\tau}^{[\theta,\phi]}(L+1,y)=\e^{i(\theta+\phi)}U_{\tau}^{[\theta,\phi]}(L,y)$. This
means that eq.~\eqref{eq:21} gives the universal form of the
time-evolution kernel for a single walker on the finite interval
subject to these boundary conditions. If one wants a theory that
satisfies the Dirichlet boundary conditions at $x=0$ and $x=L+1$, one
should redefine the translation and reflection as
$tx\coloneq x+2(L+1)$ and $rx\coloneq-x$, respectively, and choose the
representation $\phi=\pi$. In this case, one obtains
$U_{\tau}^{[\theta,\phi=\pi]}(x,y)=\sum_{-\infty}^{\infty}\e^{in\theta}[\widetilde{U}_{\tau}(x,y+2n(L+1))-\widetilde{U}_{\tau}(x,-y+2n(L+1))]$
which satisfies $U_{\tau}^{[\theta,\phi=\pi]}(0,y)=0$ and
$U_{\tau}^{[\theta,\phi=\pi]}(L+1,y)=0$.

We note in closing that eq.~\eqref{eq:21} can also be obtained from
the time-evolution kernel on a circle \eqref{eq:15} by gauging the
reflection invariance at $\theta=0,\pi\pmod{2\pi}$. In fact,
eq.~\eqref{eq:21} can be written as
$U_{\tau}^{[\theta,\phi]}(x,y)=\sum_{m=0}^{1}D^{[\phi]}(r^{m})U_{\tau}^{[\theta]}(x,r^{m}y)=\sum_{m=0}^{1}\sum_{n=-\infty}^{\infty}D^{[\phi]}(r^{m})D^{[\theta]}(t^{n})\widetilde{U}_{\tau}(x,t^{n}r^{m}y)$,
where $D^{[\phi]}$ is the one-dimensional unitary representation of
$\mathbb{Z}_{2}$ given by eq.~\eqref{eq:17}. An important lesson from
this example is that there could exist several ways to construct
time-evolution kernels on orbit spaces.

\subsection{Identical walkers in one dimension}
\label{section:3.2}
Now let us turn to the problem of multiple identical walkers on a
lattice. The key to this problem is the \textit{indistinguishability}
of identical particles, where physical observables must be invariant
under permutations of multiparticle coordinates. As is well known,
this indistinguishability always makes the multiparticle configuration
space an orbit space
\cite{Souriau:1967,Souriau:1969,Laidlaw:1970ei,Leinaas:1977fm,Harshman:2021jlv}. The
basic idea behind this is to regard the permutation invariance as a
\textit{gauge symmetry} (i.e., redundancy in description). From this
perspective, the configuration space must be a collection of
inequivalent gauge orbits because gauge-equivalent configurations are
physically equivalent.

To date, there exist two distinct formulations of this idea in
identical-particle problems. The first regards the configuration space
of $N$ identical particles as the orbit space
$(X^{N}-\Delta_{N})/S_{N}$, where $X^{N}$ is the $N$-fold Cartesian
product of a single-particle configuration space $X$ and
$\Delta_{N}\subset X^{N}$ is the set of fixed points under the action
of the symmetric group $S_{N}$
\cite{Souriau:1967,Souriau:1969,Laidlaw:1970ei,Leinaas:1977fm}. On the
other hand, the second includes the fixed points and regards the
configuration space as the orbit space $X^{N}/S_{N}$
\cite{Harshman:2021jlv}. The difference between these two formulations
is very subtle (especially in lattices) and beyond the scope of this
note. Fortunately, however, we can circumvent this issue and solve the
$N$-identical-walker problems as follows.

Suppose that $X$ itself is a nontrivial orbit space and takes the form
$X=\widetilde{X}/G$, where $G$ is a discrete group whose action on
$\widetilde{X}$ has no fixed points. In this case, the configuration
space can also be written as
$(\widetilde{X}^{N}-\widetilde{\Delta}_{N})/(G\wr S_{N})$ or
$\widetilde{X}^{N}/(G\wr S_{N})$.\footnote{Here is the proof. First,
  the wreath product $G\wr S_{N}=G^{N}\rtimes S_{N}$ can be written as
  the set $\{g\sigma:g\in G^{N},\sigma\in S_{N}\}$ equipped with the
  group composition law
  $(g\sigma)(g^{\prime}\sigma^{\prime})=(g\sigma
  g^{\prime}\sigma^{-1})(\sigma\sigma^{\prime})$ for any
  $g,g^{\prime}\in G^{N}$ and $\sigma,\sigma^{\prime}\in S_{N}$. Here
  $g\mapsto \sigma g\sigma^{-1}$ is the automorphism of the $N$-fold
  direct-product group $G^{N}=G\times\cdots\times G$ defined by
  $\sigma g\sigma^{-1}\coloneq g_{\sigma(1)}\cdots g_{\sigma(N)}$ for
  any $g=g_{1}\cdots g_{N}\in G\times\cdots\times G$. It is now
  obvious that first making the identification $x\sim gx$ by
  $g\in G^{N}$ in $\widetilde{X}^{N}$ and then making the
  identification $x\sim\sigma x$ by $\sigma\in S_{N}$ in
  $\widetilde{X}^{N}/G^{N}$ is equivalent to making the identification
  $x\sim\sigma gx$ by
  $\sigma g=(\sigma g\sigma^{-1})\sigma\in G\wr S_{N}$ in
  $\widetilde{X}^{N}$. Hence $(\widetilde{X}^{N}/G^{N})/S_{N}$ is
  equivalent to $\widetilde{X}^{N}/(G\wr S_{N})$. By subtracting the
  set of fixed points of $S_{N}$, we also see that
  $(\widetilde{X}^{N}/G^{N}-\Delta_{N})/S_{N}$ is equivalent to
  $(\widetilde{X}^{N}-\widetilde{\Delta}_{N})/(G\wr S_{N})$. See also
  refs.~\cite{Imbo:1989kj,Harshman:2021jlv} for similar results in
  continuous spaces.} Here $\wr$ stands for the wreath product defined
by the semidirect product $G\wr S_{N}\coloneq G^{N}\rtimes S_{N}$ and
$\widetilde{\Delta}_{N}\subset\widetilde{X}^{N}$ is the set of fixed
points of $S_{N}$. Hence, irrespective of the formulations, once given
a time-evolution kernel on
$\Lambda=\widetilde{X}^{N}-\widetilde{\Delta}_{N}$ or
$\widetilde{X}^{N}$, the problem just reduces to the classification of
one-dimensional unitary representations of the discrete group
$\Gamma=G\wr S_{N}$.

In this section, we shall focus on the cases $X=\mathbb{Z}$,
$\mathbb{Z}/L\mathbb{Z}$, $\mathbb{Z}/\mathbb{Z}_{2}$, and
$\mathbb{Z}/D_{\infty}$ and construct time-evolution kernels for $N$
identical walkers on the infinite line, a circle, the half line, and a
finite interval. In the following, $\widetilde{U}_{\tau}(x,y)$
represents a time-evolution kernel on
$\mathbb{Z}^{N}-\widetilde{\Delta}_{N}$ or $\mathbb{Z}^{N}$ that
satisfies the translation invariance, reflection invariance, and
permutation invariance.

\begin{figure}[t!]
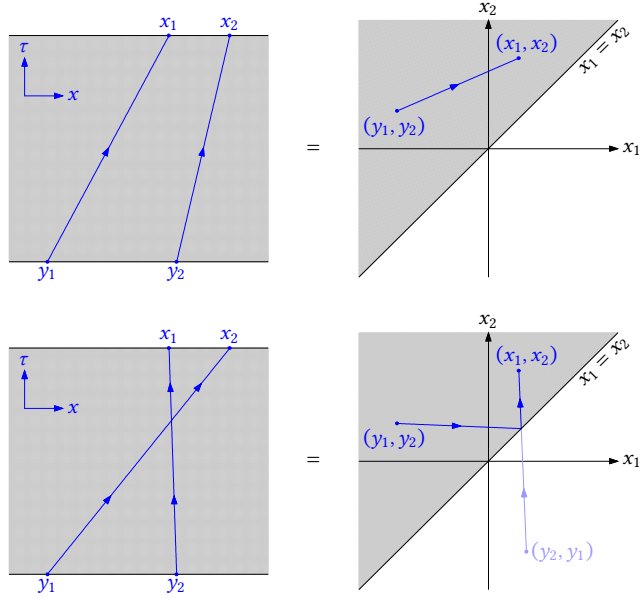

  \centering%
  \begin{tabular}{c}
    \input{figure2a.eepic}\\
    \input{figure2b.eepic}\\
  \end{tabular}
  \caption{Typical time-evolutions of two identical particles on the
    infinite line with the initial and final points $y=(y_{1},y_{2})$
    and $x=(x_{1},x_{2})$. When $N=2$, the time-evolution kernel
    \eqref{eq:25} consists of only two terms,
    $U_{\tau}^{[\pm]}(x_{1},x_{2},y_{1},y_{2})=\widetilde{U}_{\tau}(x_{1},x_{2},y_{1},y_{2})\pm
    U_{\tau}(x_{1},x_{2},y_{2},y_{1})$. In the spacetime picture, the
    first and second terms correspond to time-evolutions of
    two-identical particles without and with particle exchange,
    respectively; see the left panels. In the configuration-space
    picture, on the other hand, these terms correspond to a direct and
    reflected paths; see the right panels. In the latter picture, the
    particle-exchange phase $\pm1$ is described by the phase shift by
    reflecting off the boundary. (Note that the two-particle
    configuration space is the two-dimensional lattice with the
    identification $(x_{1},x_{2})\sim(x_{2},x_{1})$, which has the
    boundary at $x_{1}=x_{2}$ and is identical to the half space
    $(\mathbb{Z}^{2}-\Delta_{2})/S_{2}\cong\{(x_{1},x_{2})\in\mathbb{Z}^{2}:x_{1}<x_{2}\}$
    or
    $\mathbb{Z}^{2}/S_{2}\cong\{(x_{1},x_{2})\in\mathbb{Z}^{2}:x_{1}\leq
    x_{2}\}$.)}
  \label{figure:2}
\end{figure}

\paragraph*{Example 4. ($N$ identical walkers on the infinite line)}
Let us first consider $N$ identical walkers on the integer lattice
$\mathbb{Z}$. In this case, the discrete group $\Gamma=S_{N}$ is just
the symmetric group of order $N!$, whose presentation is
\begin{align}
  S_{N}=\left\langle\sigma_{1},\cdots,\sigma_{N-1}\,\middle\vert\,\sigma_{i}^{2}=e,~~\sigma_{i}\sigma_{i+1}\sigma_{i}=\sigma_{i+1}\sigma_{i}\sigma_{i+1},~~\sigma_{i}\sigma_{j}=\sigma_{j}\sigma_{i}~~(|i-j|\geq2)\right\rangle.\label{eq:22}
\end{align}
Here $\sigma_{i}=(i,i+1)$ is the adjacent transposition that
interchanges $i$ and $i+1$. Its action on
$x=(x_{1},\cdots,x_{N})\in\mathbb{Z}^{N}$ is defined as follows:
\begin{align}
  \sigma_{i}x\coloneq(x_{1},\cdots,x_{i-1},x_{i+1},x_{i},x_{i+2},\cdots,x_{N}).\label{eq:23}
\end{align}
An arbitrary element $\sigma\in S_{N}$ can be written as a product of
the generators $\sigma_{1},\cdots,\sigma_{N-1}$. Its action on
$x=(x_{1},\cdots,x_{N})$ can be written as
$\sigma x=(x_{\sigma(1)},\cdots,x_{\sigma(N)})$, where $\sigma(i)$
stands for the permutation of $i$ under $\sigma$.

Now, there exist two distinct one-dimensional unitary representations
of $S_{N}$: the trivial representation and the sign
representation. Though this result is well known, let us reproduce it
here just for later convenience. Since $\sigma_{i}^{2}=e$ and
$\sigma_{i}\sigma_{i+1}\sigma_{i}=\sigma_{i+1}\sigma_{i}\sigma_{i+1}$,
any one-dimensional unitary representation $D:S_{N}\to U(1)$ must
satisfy the conditions $D(\sigma_{i})^{2}=1$ and
$D(\sigma_{i})D(\sigma_{i+1})D(\sigma_{i})=D(\sigma_{i+1})D(\sigma_{i})D(\sigma_{i+1})$,
whose solutions are $D(\sigma_{i})=\pm1$ and
$D(\sigma_{i})=D(\sigma_{i+1})$. Hence, we have
$D(\sigma_{1})=\cdots=D(\sigma_{N-1})=\pm1$; that is, there exist two
distinct maps $D^{[\pm]}$ given by
\begin{align}
  D^{[\pm]}(\sigma)=(\pm1)^{\#\sigma},\label{eq:24}
\end{align}
where $\#\sigma$ stands for the number of adjacent transpositions in
the permutation $\sigma$. In the standard terminology, $D^{[+]}$ is
the trivial representation and $D^{[-]}$ is the sign
representation.\footnote{The sign representation can also be written
  as $D^{[-]}(\sigma)=\sgn(\sigma)$, where $\sgn(\sigma)$ stands for
  the signature of $\sigma$. It is defined by $\sgn(\sigma)=\pm1$ for
  even (odd) permutations.} Correspondingly, there exist the following
two distinct time-evolution kernels for $N$ identical walkers on
$\mathbb{Z}$:
\begin{align}
  U_{\tau}^{[\pm]}(x,y)
  &=\sum_{\sigma\in S_{N}}D^{[\pm]}(\sigma)\widetilde{U}_{\tau}(x,\sigma y)\nonumber\\
  &=\sum_{\sigma\in S_{N}}(\pm1)^{\#\sigma}\widetilde{U}_{\tau}(x,\sigma y).\label{eq:25}
\end{align}
Notice that eq.~\eqref{eq:25} satisfies the identity
$U_{\tau}^{[\pm]}(\sigma
x,y)=(\pm1)^{\#\sigma}U_{\tau}^{[\pm]}(x,y)$. The weight factors
\eqref{eq:24} thus describe particle-exchange phases under the
permutation of identical particles. It is now obvious that the two
distinct representations $D^{[\pm]}$ correspond to two distinct
particle statistics: $U_{\tau}^{[+]}$ describes the time-evolution
kernel for $N$ identical bosons, while $U_{\tau}^{[-]}$ describes that
for $N$ identical fermions. For a geometrical interpretation of
eq.~\eqref{eq:25}, see figure~\ref{figure:2}.

\paragraph*{Example 5. ($N$ identical walkers on a circle)} Let us
next consider $N$ identical particles on the periodic lattice of $L$
sites. In this case, the discrete group is the wreath product
$\Gamma=L\mathbb{Z}\wr S_{N}$, whose presentation is given by
\begin{align}
  L\mathbb{Z}\wr S_{N}=
  \left\langle
  \begin{array}{l}
    t_{1},\cdots,t_{N},\\
    \sigma_{1},\cdots,\sigma_{N-1}\\
  \end{array}
  \middle\vert
  \begin{array}{l}
    t_{i}t_{j}=t_{j}t_{i},\quad
    \sigma_{i}^{2}=e,\\
    \sigma_{i}\sigma_{i+1}\sigma_{i}=\sigma_{i+1}\sigma_{i}\sigma_{i+1},\quad
    \sigma_{i}\sigma_{j}=\sigma_{j}\sigma_{i}~~~(|i-j|\geq2),\\
    \sigma_{i}t_{i}\sigma_{i}=t_{i+1},\quad
    \sigma_{i}t_{j}\sigma_{i}=t_{j}~~~(j\neq i,i+1)
  \end{array}
  \right\rangle.\label{eq:26}
\end{align}
Here the actions of the generators $t_{i}$ and $\sigma_{i}$ are
defined by eq.~\eqref{eq:23} and
\begin{align}
  t_{i}x\coloneq(x_{1},\cdots,x_{i-1},x_{i}+L,x_{i+1},\cdots,x_{N}).\label{eq:27}
\end{align}
Note that any element of $L\mathbb{Z}\wr S_{N}$ can be written as
$t_{1}^{n_{1}}\cdots t_{N}^{n_{N}}\sigma$, where $\sigma$ is a
permutation and $n_{1},\cdots,n_{N}=0,\pm1,\cdots$. Its action on
$x=(x_{1},\cdots,x_{N})$ is given by
$t_{1}^{n_{1}}\cdots t_{N}^{n_{N}}\sigma
x=(x_{\sigma(1)}+n_{1}L,\cdots,x_{\sigma(N)}+n_{N}L)$.

Now we have to classify one-dimensional unitary representation
$D:L\mathbb{Z}\wr S_{N}\to U(1)$. First, the relations
$\sigma_{i}^{2}=e$ and
$\sigma_{i}\sigma_{i+1}\sigma_{i}=\sigma_{i+1}\sigma_{i}\sigma_{i+1}$
imply $D(\sigma_{1})=\cdots=D(\sigma_{N})=\pm1$. Second, the relation
$\sigma_{i}t_{i}\sigma_{i}=t_{i+1}$ implies
$D(\sigma_{i})D(t_{i})D(\sigma_{i})=D(t_{i+1})$, which, together with
$D(\sigma_{i})^{2}=1$, leads to
$D(t_{1})=\cdots=D(t_{N})=\e^{i\theta}$, where
$\theta\in\mathbb{R}/2\pi\mathbb{R}$. Thus we have two distinct
one-parameter families of the maps $D^{[\theta,\pm]}$ given by
\begin{align}
  D^{[\theta,\pm]}(t_{1}^{n_{1}}\cdots t_{N}^{n_{N}}\sigma)=\e^{i(n_{1}+\cdots+n_{N})\theta}(\pm1)^{\#\sigma}.\label{eq:28}
\end{align}
The time-evolution kernel for $N$ identical walkers on
$\mathbb{Z}/L\mathbb{Z}$ is therefore
\begin{align}
  U_{\tau}^{[\theta,\pm]}(x,y)=\sum_{n_{1}=-\infty}^{\infty}\cdots\sum_{n_{N}=-\infty}^{\infty}\sum_{\sigma\in S_{N}}D^{[\theta,\pm]}(t_{1}^{n_{1}}\cdots t_{N}^{n_{N}}\sigma)\widetilde{U}_{\tau}(x,t_{1}^{n_{1}}\cdots t_{N}^{n_{N}}\sigma y).\label{eq:29}
\end{align}
Notice that the kernel \eqref{eq:29} satisfies the identities
$U_{\tau}^{[\theta,\pm]}(\sigma
x,y)=(\pm1)^{\#\sigma}U_{\tau}^{[\theta,\pm]}(x,y)$ and
$U_{\tau}^{[\theta,\pm]}(t_{i}x,y)=\e^{i\theta}U_{\tau}^{[\theta,\pm]}(x,y)$
for any $i=1,\cdots,N$. Physically, $U_{\tau}^{[\theta,+]}$
($U_{\tau}^{[\theta,-]}$) describes the system of $N$ identical bosons
(fermions) on a circle with a nonzero magnetic flux.

\paragraph*{Example 6. ($N$ identical walkers on the half line)} Let
us next consider $N$ identical particles on the semi-infinite
lattice. In this case, the discrete group is
$\Gamma=\mathbb{Z}_{2}\wr S_{N}$, where
\begin{align}
  \mathbb{Z}_{2}\wr S_{N}=
  \left\langle
  \begin{array}{l}
    r_{1},\cdots,r_{N},\\
    \sigma_{1},\cdots,\sigma_{N-1}\\
  \end{array}
  \middle\vert
  \begin{array}{l}
    r_{i}r_{j}=r_{j}r_{i},\quad
    r_{i}^{2}=\sigma_{i}^{2}=e,\\
    \sigma_{i}\sigma_{i+1}\sigma_{i}=\sigma_{i+1}\sigma_{i}\sigma_{i+1},\quad
    \sigma_{i}\sigma_{j}=\sigma_{j}\sigma_{i}~~~(|i-j|\geq2),\\
    \sigma_{i}r_{i}\sigma_{i}=r_{i+1},\quad
    \sigma_{i}r_{j}\sigma_{i}=r_{j}~~~(j\neq i,i+1)
  \end{array}
  \right\rangle.\label{eq:30}
\end{align}
The actions of the generators are defined by eq.~\eqref{eq:23} and
\begin{align}
  r_{i}x\coloneq(x_{1},\cdots,x_{i-1},1-x_{i},x_{i+1},\cdots,x_{N}).\label{eq:31}
\end{align}
Note that any element of $\mathbb{Z}_{2}\wr S_{N}$ can be written as
the product $r_{1}^{n_{1}}\cdots r_{N}^{n_{N}}\sigma$, where
$\sigma\in S_{N}$ and $n_{1},\cdots,n_{N}=0,1$. Its action on
$x=(x_{1},\cdots,x_{N})$ is given by
$r_{1}^{n_{1}}\cdots r_{N}^{n_{N}}\sigma
x=(\cdots,x_{\sigma(i)},\cdots)$ for $n_{i}=0$ and
$r_{1}^{n_{1}}\cdots r_{N}^{n_{N}}\sigma
x=(\cdots,1-x_{\sigma(i)},\cdots)$ for $n_{i}=1$.

By repeating the same procedure as above, one can show that
one-dimensional unitary representation
$D:\mathbb{Z}_{2}\wr S_{N}\to U(1)$ must satisfy
$D(r_{1})=\cdots=D(r_{N})=\pm1$ and
$D(\sigma_{1})=\cdots=D(\sigma_{N-1})=\pm1$. Hence there exist
$2^{2}=4$ distinct maps $D^{[\phi,\pm]}$ given by
\begin{align}
  D^{[\phi,\pm]}(r_{1}^{n_{1}}\cdots r_{N}^{n_{N}}\sigma)=\e^{i(n_{1}+\cdots+n_{N})\phi}(\pm1)^{\#\sigma},\label{eq:32}
\end{align}
where $\phi\in\{0,\pi\pmod{2\pi}\}$. The time-evolution kernel for $N$
identical walkers on $\mathbb{Z}/\mathbb{Z}_{2}$ is therefore
\begin{align}
  U_{\tau}^{[\phi,\pm]}(x,y)=\sum_{n_{1}=0}^{1}\cdots\sum_{n_{N}=0}^{1}\sum_{\sigma\in S_{N}}D^{[\phi,\pm]}(r_{1}^{n_{1}}\cdots r_{N}^{n_{N}}\sigma)\widetilde{U}_{\tau}(x,r_{1}^{n_{1}}\cdots r_{N}^{n_{N}}\sigma y).\label{eq:33}
\end{align}
Notice that eq.~\eqref{eq:33} satisfies
$U_{\tau}^{[\phi,\pm]}(\sigma
x,y)=(\pm1)^{\#\sigma}U_{\tau}^{[\phi,\pm]}(x,y)$ and
$U_{\tau}^{[\phi,\pm]}(r_{i}x,y)=\e^{i\phi}U_{\tau}^{[\phi,\pm]}(x,y)$
for any $i=1,\cdots,N$. Hence, $U_{\tau}^{[\phi,\pm]}$ describes the
system of $N$ identical bosons (fermions) that acquire the phase shift
$\phi$ when reflected off the boundary.

\paragraph*{Example 7. ($N$ identical walkers on a finite interval)}
Let us finally consider $N$ identical particles on a finite
interval. In this case, the discrete group is
$\Gamma=D_{\infty}\wr S_{N}$, where
\begin{align}
  D_{\infty}\wr S_{N}=
  \left\langle
  \begin{array}{l}
    t_{1},\cdots,t_{N},\\
    r_{1},\cdots,r_{N},\\
    \sigma_{1},\cdots,\sigma_{N-1}\\
  \end{array}
  \middle\vert
  \begin{array}{l}
    t_{i}t_{j}=t_{j}t_{i},\quad
    r_{i}r_{j}=r_{j}r_{i},\quad
    r_{i}^{2}=\sigma_{i}^{2}=e,\\
    r_{i}t_{i}r_{i}=t_{i}^{-1},\quad
    r_{i}t_{j}r_{i}=t_{j}~~~(j\neq i),\\
    \sigma_{i}\sigma_{i+1}\sigma_{i}=\sigma_{i+1}\sigma_{i}\sigma_{i+1},\quad
    \sigma_{i}\sigma_{j}=\sigma_{j}\sigma_{i}~~~(|i-j|\geq2),\\
    \sigma_{i}r_{i}\sigma_{i}=r_{i+1},\quad
    \sigma_{i}r_{j}\sigma_{i}=r_{j}~~~(j\neq i,i+1)
  \end{array}
  \right\rangle.\label{eq:34}
\end{align}
The actions of the generators are given by eqs.~\eqref{eq:23},
\eqref{eq:31}, and
\begin{align}
  t_{i}x\coloneq(x_{1},\cdots,x_{i-1},x_{i}+2L,x_{i+1},\cdots,x_{N}).\label{eq:35}
\end{align}
We note that any element of $D_{\infty}\wr S_{N}$ can be written as
the product
$t_{1}^{n_{1}}r_{1}^{m_{1}}\cdots t_{N}^{n_{N}}r_{N}^{m_{N}}\sigma$,
where $\sigma\in S_{N}$, $n_{1},\cdots,n_{N}=0,\pm1,\pm2,\cdots$, and
$m_{1},\cdots,m_{N}=0,1$. Its action is given by
$t_{1}^{n_{1}}r_{1}^{m_{1}}\cdots t_{N}^{n_{N}}r_{N}^{m_{N}}\sigma
x=(\cdots,x_{\sigma(i)}+2n_{i}L,\cdots)$ for $m_{i}=0$ and
$t_{1}^{n_{1}}r_{1}^{m_{1}}\cdots t_{N}^{n_{N}}r_{N}^{m_{N}}\sigma
x=(\cdots,1-x_{\sigma(i)}+2n_{i}L,\cdots)$ for $m_{i}=1$.

Now it is a straightforward exercise to show that there exist
$2^{3}=8$ distinct one-dimensional unitary representations of the
wreath product $D_{\infty}\wr S_{N}$. The result is the following
maps:
\begin{align}
  D^{[\theta,\phi,\pm]}(t_{1}^{n_{1}}r_{1}^{m_{1}}\cdots t_{N}^{n_{N}}r_{N}^{m_{N}}\sigma)=\e^{i(n_{1}+\cdots+n_{N})\theta}\e^{i(m_{1}+\cdots+m_{N})\phi}(\pm1)^{\#\sigma},\label{eq:36}
\end{align}
where $\theta,\phi\in\{0,\pi\pmod{2\pi}\}$. Correspondingly, we have
the following eight distinct time-evolution kernels for $N$ identical
walkers on $\mathbb{Z}/D_{\infty}$:
\begin{align}
  U_{\tau}^{[\theta,\phi,\pm]}(x,y)=\sum_{n_{1}=-\infty}^{\infty}\sum_{m_{1}=0}^{1}\cdots\sum_{n_{N}=-\infty}^{\infty}\sum_{m_{N}=0}^{1}\sum_{\sigma\in S_{N}}D^{[\theta,\phi,\pm]}(t_{1}^{n_{1}}r_{1}^{m_{1}}\cdots t_{N}^{n_{N}}r_{N}^{m_{N}}\sigma)\widetilde{U}_{\tau}(x,t_{1}^{n_{1}}r_{1}^{m_{1}}\cdots t_{N}^{n_{N}}r_{N}^{m_{N}}\sigma y).\label{eq:37}
\end{align}
Physically, $U_{\tau}^{[\theta,\phi,\pm]}$ describes the system of $N$
identical bosons (fermions) that acquire the phase shifts $\phi$ and
$\theta+\phi$ when reflected off the boundaries $x=1$ and $x=L$,
respectively.

\section{Asides}
\label{section:4}
Now, there exist several other quantities that can be constructed in
exactly the same way as for the time-evolution kernel
\eqref{eq:4}. Examples include the resolvent kernel (Green's function)
and the canonical density matrix (density matrix in the canonical
ensemble). Another example is a unitary representation of an arbitrary
group $G$ on a (tensor-product) Hilbert space, which includes the
time-evolution kernel in discrete-time quantum walk. In this section,
we shall briefly discuss the construction of these quantities on the
orbit space $\Lambda/\Gamma$.

\subsection{Resolvent kernel}
\label{section:4.1}
Let us first start with the resolvent kernel---a matrix element of the
resolvent operator in position space. Let $H$ be the Hamiltonian
operator of the system. Then, the resolvent operator
$G_{E}=(EI-H)^{-1}$ for $\im E>0$ and the time-evolution operator
$U_{\tau}=\e^{-iH\tau}$ for $\tau>0$ are transformed into one another
through the Laplace transform
$i(EI-H)^{-1}=\int_{0}^{\infty}\!d\tau\e^{-iH\tau}\e^{iE\tau}$ and the
inverse Laplace transform
$\e^{-iH\tau}=\int_{-\infty+i\epsilon}^{\infty+i\epsilon}\!\frac{dE}{2\pi}\,i(EI-H)^{-1}\e^{-iE\tau}$,
respectively, where $\epsilon$ is an arbitrary positive
real. Consequently, the matrix elements
$U_{\tau}(x,y)=\langle x|U_{\tau}|y\rangle$ and
$G_{E}(x,y)=\langle x|G_{E}|y\rangle$ are mutually related through the
following:
\begin{subequations}
  \begin{alignat}{2}
    iG_{E}(x,y)&=\int_{0}^{\infty}\!\!d\tau\,U_{\tau}(x,y)\e^{iE\tau}&&\quad\text{for}\quad\im E>0,\label{eq:38a}\\
    U_{\tau}(x,y)&=\int_{-\infty+i\epsilon}^{\infty+i\epsilon}\!\frac{dE}{2\pi}\,iG_{E}(x,y)\e^{-iE\tau}&&\quad\text{for}\quad \tau>0.\label{eq:38b}
  \end{alignat}
\end{subequations}
Hence, by applying the Laplace transform to formula \eqref{eq:4}, we
find that the resolvent kernel on $\Lambda/\Gamma$ takes the following
form:
\begin{align}
  G_{E}(x,y)=\sum_{\gamma\in\Gamma}D(\gamma)\widetilde{G}_{E}(x,\gamma y),\label{eq:39}
\end{align}
where
$i\widetilde{G}_{E}(x,y)=\int_{0}^{\infty}\!d\tau\,\widetilde{U}_{\tau}(x,y)\e^{iE\tau}$
($\im E>0$) is the resolvent kernel on $\Lambda$.

An immediate application of the above formula is the local density of
states given by $\rho_{E}(x)=\langle x|\delta(EI-H)|x\rangle$. In
fact, by using the identity
\begin{align}
  \lim_{\im E\to0_{+}}(EI-H)^{-1}=\mathscr{P}(EI-H)^{-1}-i\pi\delta(EI-H),\label{eq:40}
\end{align}
where $\mathscr{P}$ stands for the Cauchy principal value, we find
$\im G_{E}(x,x)=\im\langle x|(EI-H)^{-1}|x\rangle=-\pi\langle
x|\delta(EI-H)|x\rangle=-\pi\rho_{E}(x)$ in the limit $\im
E\to0_{+}$. Thus,
\begin{align}
  \rho_{E}(x)=-\frac{1}{\pi}\im\sum_{\gamma\in\Gamma}D(\gamma)\widetilde{G}_{E}(x,\gamma x)\quad\text{as}\quad\im E\to0_{+}.\label{eq:41}
\end{align}
The density of states $\rho_{E}=\tr\delta(EI-H)$ then takes the form
$\rho_{E}=-(1/\pi)\im\sum_{x\in\Lambda/\Gamma}\sum_{\gamma\in\Gamma}D(\gamma)\widetilde{G}_{E}(x,\gamma
x)$.

\subsection{Canonical density matrix}
\label{section:4.2}
Let us next consider the canonical density matrix on
$\Lambda/\Gamma$. In thermal equilibrium at temperature $\beta^{-1}$,
the canonical density matrix is given by
$\rho_{\beta}=U_{-i\beta}/Z(\beta)$, where $U_{-i\beta}=\e^{-\beta H}$
is the Gibbs operator and $Z(\beta)=\tr U_{-i\beta}$ is the canonical
partition function. Note that the Gibbs operator satisfies the
composition law
$U_{-i\beta_{1}}U_{-i\beta_{2}}=U_{-i(\beta_{1}+\beta_{2})}$, the
hermiticity $U_{-i\beta}^{\dagger}=U_{-i\beta}$, and the initial
condition $U_{0}=I$. Its matrix elements (heat kernel)
$U_{-i\beta}(x,y)=\langle x|\e^{-\beta H}|y\rangle$ must then satisfy
these conditions as well. Namely, we must have
$\sum_{z\in\Lambda/\Gamma}U_{-i\beta_{1}}(x,z)U_{-i\beta_{2}}(z,y)=U_{-i(\beta_{1}+\beta_{2})}(x,y)$,
$\overline{U_{-i\beta}(x,y)}=U_{-i\beta}(y,x)$, and
$U_{0}(x,y)=\delta_{x,y}$. Under these conditions, one can again show
that $U_{-i\beta}(x,y)$ can be written as
$U_{-i\beta}(x,y)=\sum_{\gamma\in\Gamma}D(\gamma)\widetilde{U}_{-i\beta}(x,\gamma
y)$. Hence the matrix elements of the canonical density matrix are
\begin{align}
  \rho_{\beta}(x,y)=\frac{1}{Z(\beta)}\sum_{\gamma\in\Gamma}D(\gamma)\widetilde{U}_{-i\beta}(x,\gamma y),\label{eq:42}
\end{align}
where $\rho_{\beta}(x,y)=\langle x|\rho_{\beta}|y\rangle$. Here
$Z(\beta)$ is the canonical partition function given by
\begin{align}
  Z(\beta)=\sum_{x\in\Lambda/\Gamma}\sum_{\gamma\in\Gamma}D(\gamma)\widetilde{U}_{-i\beta}(x,\gamma x).\label{eq:43}
\end{align}
We note that the partition function \eqref{eq:43} can also be written
as
$Z(\beta)=\sum_{x\in\Lambda/\Gamma}\sum_{\gamma\in\Gamma}D(\gamma)\langle
x|\e^{-\beta\widetilde{H}}|\gamma
x\rangle=\sum_{\gamma\in\Gamma}D(\gamma)\tr(\e^{-\beta\widetilde{H}}W_{\gamma})$,
where $\widetilde{H}$ is the Hamiltonian operator on $\Lambda$ and
$W_{\gamma}$ is a unitary operator defined by
$W_{\gamma}|x\rangle=|\gamma x\rangle$.

\subsection{Unitary representations of arbitrary groups on a
  tensor-product Hilbert space}
\label{section:4.3}
As mentioned in the beginning of section~\ref{section:2}, our main
formula \eqref{eq:4} is also applicable to discrete-time quantum walk,
where the time $\tau$ takes discrete values and the one-particle
Hilbert space is the tensor product of the position and coin Hilbert
spaces. In this section, we shall see this from a more general
perspective: the construction of matrix elements of a unitary
representation of an arbitrary group $G$ on a tensor-product Hilbert
space. The time-evolution kernel in discrete-time quantum walk just
corresponds to the special case $G=\mathbb{Z}$ (the additive group of
integers).

To begin with, let $\{U_{g}\in U(\mathscr{H}):g\in G\}$ be a unitary
representation of $G$ on the tensor-product Hilbert space
$\mathscr{H}=\mathscr{H}_{\text{position}}\otimes\mathscr{H}_{\text{coin}}$,
where $U(\mathscr{H})$ stands for the set of unitary operators on
$\mathscr{H}$, $\mathscr{H}_{\text{position}}=l^{2}(\Lambda/\Gamma)$
is the set of square-summable sequences on the orbit space
$\Lambda/\Gamma$, and $\mathscr{H}_{\text{coin}}=\mathbb{C}^{d}$ is
the $d$-dimensional complex vector space that describes internal
degrees of freedom of particles. Let $\{|x\rangle\}$ and
$\{|i\rangle\}$ be complete orthonormal systems of
$\mathscr{H}_{\text{position}}$ and $\mathscr{H}_{\text{coin}}$,
respectively. The set $\{|x\rangle\otimes|i\rangle\}$ then provides a
complete orthonormal system of the total Hilbert space $\mathscr{H}$
such that the matrix elements of $U_{g}$ can be defined as
$U_{g}(x,i;y,j)=(\langle x|\otimes\langle
i|)U_{g}(|y\rangle\otimes|j\rangle)$.

We now define $U_{g}(x,y)$ as the following $d\times d$ matrix:
\begin{align}
  U_{g}(x,y)\coloneq
  \begin{pmatrix}
    U_{g}(x,1;y,1)&\cdots&U_{g}(x,1;y,d)\\
    \vdots&\ddots&\vdots\\
    U_{g}(x,d;y,1)&\cdots&U_{g}(x,d;y,d)\\
  \end{pmatrix}.\label{eq:44}
\end{align}
Since the unitary representation must satisfy the group composition
law $U_{g_{1}}U_{g_{2}}=U_{g_{1}g_{2}}$, the unitarity
$U_{g}^{\dagger}(=U_{g}^{-1})=U_{g^{-1}}$, and the initial condition
$U_{e}=I$, matrix \eqref{eq:44} must also satisfy the following
properties:
\begin{subequations}
  \begin{align}
    \sum_{z\in\Lambda/\Gamma}U_{g_{1}}(x,z)U_{g_{2}}(z,y)&=U_{g_{1}g_{2}}(x,y),\label{eq:45a}\\
    \overline{{}^{t}U_{g}(x,y)}&=U_{g^{-1}}(y,x),\label{eq:45b}\\
    U_{e}(x,y)&=\delta_{x,y}\bm{1},\label{eq:45c}
  \end{align}
\end{subequations}
where $x,y\in\Lambda/\Gamma$. Here $t$ and $\bm{1}$ stand for the
matrix transpose and the $d\times d$ identity matrix,
respectively. Now it is a straightforward exercise to show that matrix
\eqref{eq:44} can be written as
\begin{align}
  U_{g}(x,y)=\sum_{\gamma\in\Gamma}D(\gamma)\widetilde{U}_{g}(x,\gamma y),\label{eq:46}
\end{align}
where $\widetilde{U}_{g}(x,y)$ is a $d\times d$ matrix subject to the
conditions
$\sum_{z\in\Lambda}\widetilde{U}_{g_{1}}(x,z)\widetilde{U}_{g_{2}}(z,y)=\widetilde{U}_{g_{1}g_{2}}(x,y)$,
$\overline{{}^{t}\widetilde{U}_{g}(x,y)}=\widetilde{U}_{g^{-1}}(y,x)$,
$\widetilde{U}_{e}(x,y)=\delta_{x,y}\bm{1}$, and
$\widetilde{U}_{g}(\gamma x,\gamma y)=\widetilde{U}_{g}(x,y)$ for any
$x,y\in\Lambda$ and $\gamma\in\Gamma$. It is also straightforward to
show that eq.~\eqref{eq:46} satisfies the following boundary
condition:
\begin{align}
  U_{g}(\gamma x,y)=D(\gamma)U_{g}(x,y),\quad\forall\gamma\in\Gamma.\label{eq:47}
\end{align}
It is now obvious that eq.~\eqref{eq:46} provides the time-evolution
kernel of continuous-time quantum walk with internal degrees of
freedom when $G=\mathbb{R}$ (the additive group of real numbers) and
of discrete-time quantum walk when $G=\mathbb{Z}$ (the additive group
of integers). It is also obvious that the examples presented in
section~\ref{section:3} apply to discrete-time quantum walk as well.

\section{Conclusion}
\label{section:5}
Inspired by the covering-space method in path integral on multiply
connected spaces, we have developed a general theory of quantum walk
on orbit spaces. In this note, we have proved the universal formulas
for time-evolution kernels, resolvent kernels, canonical density
matrices, and unitary representations of arbitrary groups in
continuous- and discrete-time quantum walks on the orbit space
$\Lambda/\Gamma$, where $\Lambda$ is an arbitrary lattice and $\Gamma$
is a discrete group whose action on $\Lambda$ has no fixed points. All
of these quantities are given by summations over the orbit of initial
point on $\Lambda$, where each orbit is weighted by a phase factor
given by a one-dimensional unitary representations of $\Lambda$.

There are several advantages of this orbit-space method. A main
advantage is its universality: our formulas are just based on
geometric and group-theoretic structures of configuration spaces so
that they are robust against any perturbations or interparticle
interactions as long as boundary conditions \eqref{eq:8} remain
unchanged. Another advantage is its computational simplicity: in our
formalism, one just needs to compute matrix elements on $\Lambda$,
which is generally much easier than computations on $\Lambda/\Gamma$.

Finally, let us comment on one possible future direction of this
work. A promising direction would be a generalization of our formulas
to the problem of identical walkers on graphs. Recent studies have
shown that exotic statistics may show up in many-body problems of
identical particles on graphs
\cite{Harrison:2010,Harrison:2013,Maciazek:2018,An:2020qyl,Maciazek:2020yee}. Such
exotic statistics are generalizations of braid-group statistics in two
dimensions. Hence, just as in topological quantum computation using
anyons \cite{Kitaev:1997wr}, they would have potential applications in
quantum computer science. Our formalism and its generalization may
well serve as a basic tool for studying the dynamics as well as
thermodynamics of such systems.

\subsection*{Acknowledgment}
The author would like to thank Naoto Namekata for discussion.

\appendix%
\setcounter{equation}{0}
\renewcommand{\theequation}{\thesection.\arabic{equation}}
\section{Sample computations}
\label{appendix:A}
Continuous-time quantum walk is just equivalent to tight-binding
models in condensed matter physics. The advantage of this perspective
is that it is straightforward to study many-particle problems by using
the second-quantization formalism. In this section, we study
tight-binding models for free spinless particles in one dimension and
present sample computations that justify the formulas in
section~\ref{section:3}.

\subsection{Tight-binding model on the infinite line}
\label{appendix:A.1}
Let us first consider spinless particles on the integer lattice
$\mathbb{Z}$ only with a nearest-neighbor coupling. In the
second-quantization formalism, the Hamiltonian operator is given by
\begin{align}
  \widetilde{H}=-\frac{\omega}{2}\sum_{x\in\mathbb{Z}}\left(a_{x+1}^{\dagger}a_{x}+a_{x}^{\dagger}a_{x+1}\right),\label{eq:A.1}
\end{align}
where $\omega(>0)$ is a hopping parameter. $a_{x}$ and
$a_{x}^{\dagger}$ are annihilation and creation operators for spinless
bosons (fermions) and subject to the following (anti-)commutation
relations:
\begin{align} [a_{x},a_{y}^{\dagger}]_{\mp}=\delta_{x,y}
  \quad\text{and}\quad [a_{x},a_{y}]_{\mp}=0,\label{eq:A.2}
\end{align}
where $[A,B]_{\mp}=AB\mp BA$.

Let $|0\rangle$ be the Fock vacuum that satisfies $a_{x}|0\rangle=0$
for all $x$. The time-evolution kernel in the one-particle sector of
the model is then given by
\begin{align}
  \widetilde{U}_{\tau}(x,y)=\langle x|\e^{-i\widetilde{H}\tau}|y\rangle,\label{eq:A.3}
\end{align}
where $|x\rangle=a_{x}^{\dagger}|0\rangle$ is the position-space basis
in the one-particle sector. It satisfies the orthonormality
$\langle
x|y\rangle=\langle0|a_{x}a_{y}^{\dagger}|0\rangle=\delta_{x,y}$ for
both bosons and fermions. In order to calculate the matrix element
\eqref{eq:A.3}, we first diagonalize the Hamiltonian operator, which
can be achieved by the following Fourier integral:
\begin{align}
  a_{x}=\int_{-\pi}^{\pi}\!\frac{dp}{2\pi}\,\widetilde{a}_{p}\e^{ipx},\label{eq:A.4}
\end{align}
where $\widetilde{a}_{p}$ and $\widetilde{a}_{p}^{\dagger}$ satisfy
$[\widetilde{a}_{p},\widetilde{a}_{q}^{\dagger}]_{\mp}=2\pi\delta(p-q)$
and $[\widetilde{a}_{p},\widetilde{a}_{q}]_{\mp}=0$ for any
$p,q\in(-\pi,\pi)$. By substituting eq.~\eqref{eq:A.4} into
eq.~\eqref{eq:A.3}, we obtain
\begin{align}
  \widetilde{H}=\int_{-\pi}^{\pi}\!\frac{dp}{2\pi}\,\varepsilon_{p}\widetilde{a}_{p}^{\dagger}\widetilde{a}_{p},\label{eq:A.5}
\end{align}
where $\varepsilon_{p}=-\omega\cos(p)$ is the single-particle energy
eigenvalue. It is now easy to see that the time-evolution kernel
\eqref{eq:A.3} takes the following form:
\begin{align}
  \widetilde{U}_{\tau}(x,y)
  &=\langle0|a_{x}\e^{-i\widetilde{H}\tau}a_{y}^{\dagger}|0\rangle\nonumber\\
  &=\int_{-\pi}^{\pi}\!\frac{dp}{2\pi}\int_{-\pi}^{\pi}\!\frac{dq}{2\pi}\,\langle0|\widetilde{a}_{p}\e^{-i\widetilde{H}\tau}\widetilde{a}_{q}^{\dagger}|0\rangle\e^{ipx-iqy}\nonumber\\
  &=\int_{-\pi}^{\pi}\!\frac{dp}{2\pi}\int_{-\pi}^{\pi}\!\frac{dq}{2\pi}\,\e^{-i\varepsilon_{q}t}\langle0|\widetilde{a}_{p}\widetilde{a}_{q}^{\dagger}|0\rangle\e^{ipx-iqy}\nonumber\\
  &=\int_{-\pi}^{\pi}\!\frac{dp}{2\pi}\e^{i\omega\tau\cos(p)}\e^{ip(x-y)},\label{eq:A.6}
\end{align}
where in the third equality we have used
$\e^{-iH\tau}\widetilde{a}_{q}^{\dagger}|0\rangle=\e^{-i\varepsilon_{q}t}\widetilde{a}_{q}^{\dagger}|0\rangle$,
which follows from
$\e^{-iH\tau}\widetilde{a}_{q}^{\dagger}\e^{iH\tau}=\e^{-i\varepsilon_{q}t}\widetilde{a}_{q}^{\dagger}$
and $\e^{-iH\tau}|0\rangle=|0\rangle$ (or, equivalently,
$[H,\widetilde{a}_{q}^{\dagger}]=\varepsilon_{q}\widetilde{a}_{q}^{\dagger}$
and $H|0\rangle=0$). The fourth equality follows from
$\langle0|\widetilde{a}_{p}\widetilde{a}_{q}^{\dagger}|0\rangle=2\pi\delta(p-q)$
for both bosons and fermions. To evaluate the last integral in
eq.~\eqref{eq:A.6}, we note that $\e^{i\omega\tau\cos(p)}$ is a
generating function of the Bessel function of the first kind
$J_{n}$. In fact,
\begin{align}
  \e^{i\omega\tau\cos(p)}
  &=\sum_{n=-\infty}^{\infty}\e^{in(p+\frac{\pi}{2})}J_{n}(\omega\tau)\nonumber\\
  &=\sum_{n=-\infty}^{\infty}\e^{i\frac{\pi}{2}|n|}J_{|n|}(\omega\tau)\e^{-inp},\label{eq:A.7}
\end{align}
where the second equality follows from
$J_{-n}(x)=\e^{in\pi}J_{n}(x)$. By substituting eq.~\eqref{eq:A.7}
into eq.~\eqref{eq:A.6} and then using the orthogonal relation
$\int_{-\pi}^{\pi}\!\frac{dp}{2\pi}\e^{ip(x-y-n)}=\delta_{n,x-y}$, we
obtain
\begin{align}
  \widetilde{U}_{\tau}(x,y)=\e^{i\frac{\pi}{2}|x-y|}J_{|x-y|}(\omega\tau),\quad\forall x,y\in\mathbb{Z}.\label{eq:A.8}
\end{align}
This is the well-known transition amplitude for a single walker on the
lattice $\mathbb{Z}$ (see, e.g., ref.~\cite{Farhi:1991dy}). Note that
eq.~\eqref{eq:A.8} satisfies the composition law \eqref{eq:3a}, the
unitarity \eqref{eq:3b}, and the initial condition \eqref{eq:3c},
which follow from the addition theorem
$J_{n_{1}-n_{2}}(x_{1}+x_{2})\e^{i\frac{\pi}{2}(n_{1}-n_{2})}=\sum_{n\in\mathbb{Z}}J_{n_{1}-n}(x_{1})J_{n_{2}-n}(x_{2})\e^{i\frac{\pi}{2}(n_{1}-n)}\e^{i\frac{\pi}{2}(n_{2}-n)}$
($n_{1},n_{2}\in\mathbb{Z}$), the analytic continuation
$J_{n}(\e^{i\pi}x)=\e^{in\pi}J_{n}(x)$, and $J_{n}(0)=\delta_{n,0}$,
respectively. Note also that eq.~\eqref{eq:A.8} enjoys the translation
invariance $\widetilde{U}_{\tau}(x+z,y+z)=\widetilde{U}_{\tau}(x,y)$
and the reflection invariance
$\widetilde{U}_{\tau}(z-x,z-y)=\widetilde{U}_{\tau}(x,y)$ for any
$x,y,z\in\mathbb{Z}$. As we shall see shortly, eq.~\eqref{eq:A.8}
provides the building block for the construction of time-evolution
kernels for a free particle on a circle, the half line, and a finite
interval.

Several comments are in order.
\begin{itemize}
\item\textbf{Resolvent kernel for a single walker.} As discussed in
  section~\ref{section:4.1}, the resolvent kernel (Green's function)
  is given by the Laplace transform of
  $\widetilde{U}_{\tau}(x,y)$. Let $E$ be a complex number with
  $\im E>0$. Then we have
  \begin{align}
    i\widetilde{G}_{E}(x,y)
    &=\int_{0}^{\infty}\!\!d\tau\,\widetilde{U}_{\tau}(x,y)\e^{iE\tau}\nonumber\\
    &=\int_{-\pi}^{\pi}\!\frac{dp}{2\pi}\,\frac{i\e^{ip(x-y)}}{E+\omega\cos(p)}\nonumber\\
    &=\frac{2i}{\omega}\oint_{\!\!|z|=1}\frac{dz}{2\pi i}\,\frac{z^{|x-y|}}{z^{2}+\frac{2E}{\omega}z+1},\label{eq:A.9}
  \end{align}
  where in the second equality we have substituted the last line of
  eq.~\eqref{eq:A.6} and performed the integration with respect to
  $\tau$. In the last equality we have changed the integration
  variable from $p$ to $z=\e^{ip}$, where the integration is over the
  closed loop $|z|=1$ in the counter clockwise direction. By using the
  residue theorem we find
  \begin{align}
    i\widetilde{G}_{E}(x,y)=\frac{\e^{ip|x-y|}}{\omega\sin(p)},\label{eq:A.10}
  \end{align}
  where we have parameterized the energy as $E=-\omega\cos(p)$ with
  $\re p\in(0,\pi)$ and $\im p\in(0,\infty)$. Eq.~\eqref{eq:A.10}
  provides the building block for the construction of single-particle
  resolvent kernels on a circle, the half line, and a finite interval.
\item\textbf{Heat kernel for a single walker.} The matrix element of
  the Gibbs operator $\e^{-\beta\widetilde{H}}$ can be calculated in
  exactly the same way as for $\widetilde{U}_{\tau}(x,y)$. Under the
  substitution $\tau\to-i\beta$ in eq.~\eqref{eq:A.6} we find
  \begin{align}
    \widetilde{U}_{-i\beta}(x,y)
    &=\int_{-\pi}^{\pi}\!\frac{dp}{2\pi}\e^{\beta\omega\cos(p)}\e^{ip(x-y)}\nonumber\\
    &=I_{x-y}(\beta\omega),\label{eq:A.11}
  \end{align}
  where $I_{n}(x)=I_{-n}(x)$ stands for the modified Bessel function
  of the first kind. Here in the last line we have used the fact that
  $\e^{\beta\omega\cos(p)}$ is the generating function of
  $I_{n}(\beta\omega)$. In fact,
  \begin{align}
    \e^{\beta\omega\cos(p)}=\sum_{n=-\infty}^{\infty}I_{n}(\beta\omega)\e^{-inp}.\label{eq:A.12}
  \end{align}
  By substituting this into the first line and using the orthogonal
  relation
  $\int_{-\pi}^{\pi}\!\frac{dp}{2\pi}\e^{ip(x-y-n)}=\delta_{n,x-y}$,
  we arrive at eq.~\eqref{eq:A.11}. As discussed in
  section~\ref{section:4.2}, eq.~\eqref{eq:A.11} provides the building
  block for the construction of canonical density matrices for free
  particles on a circle and a finite interval.
\item\textbf{Time-evolution kernel for $N$ identical walkers.} In the
  second-quantization formalism, it is easy to generalize the above
  results to many-particle problems. First, the position-space basis
  in the $N$-particle sector is given by
  \begin{align}
    |x_{1},\cdots,x_{N}\rangle\coloneq a_{x_{1}}^{\dagger}\cdots a_{x_{N}}^{\dagger}|0\rangle.\label{eq:A.13}
  \end{align}
  Notice that eq.~\eqref{eq:A.13} satisfies the orthonormality
  condition on the orbit space
  $(\mathbb{Z}^{N}-\Delta_{N})/S_{N}\cong\{(x_{1},\cdots,x_{N})\in\mathbb{Z}^{N}:x_{1}<\cdots<x_{N}\}$. In
  fact, for $x_{1}<\cdots<x_{N}$ and $y_{1}<\cdots<y_{N}$, we have
  \begin{align}
    \langle x_{1},\cdots,x_{N}|y_{1},\cdots,y_{N}\rangle
    &=\langle0|a_{x_{N}}\cdots a_{x_{1}}a_{y_{1}}^{\dagger}\cdots a_{y_{N}}^{\dagger}|0\rangle\nonumber\\
    &=\sum_{\sigma\in S_{N}}(\pm1)^{\#\sigma}\delta_{x_{\sigma(1)},y_{1}}\cdots\delta_{x_{\sigma(N)},y_{N}}\nonumber\\
    &=\delta_{x_{1},y_{1}}\cdots\delta_{x_{N},y_{N}},\label{eq:A.14}
  \end{align}
  where the last line follows from the fact that
  $(x_{\sigma(1)},\cdots,x_{\sigma(N)})$ and $(y_{1},\cdots,y_{N})$
  cannot be equal except for $\sigma=e$. It is now easy to show that
  the time-evolution kernels for $N$ identical bosons and fermions
  take the following forms:\footnote{It should be noted that
    $\widetilde{U}_{\tau}(x_{1},\cdots,x_{N},y_{1},\cdots,y_{N})=\prod_{j=1}^{N}\e^{i\frac{\pi}{2}|x_{j}-y_{j}|}J_{|x_{j}-y_{j}|}(\omega\tau)$
    is equivalent to a single-particle time-evolution kernel on
    $\mathbb{Z}^{N}$ rather than $\mathbb{Z}^{N}-\Delta_{N}$. As noted
    in the beginning of section~\ref{section:3.2}, in this note we
    will not touch upon this type of issues related to the fixed
    points of $S_{N}$.}
  \begin{align}
    \langle x_{1},\cdots,x_{N}|\e^{-i\widetilde{H}\tau}|y_{1},\cdots,y_{N}\rangle
    &=\langle0|a_{x_{N}}\cdots a_{x_{1}}\e^{-i\widetilde{H}\tau}a_{y_{1}}^{\dagger}\cdots a_{y_{N}}^{\dagger}|0\rangle\nonumber\\
    &=\left[\prod_{j=1}^{N}\int_{-\pi}^{\pi}\!\frac{dp_{j}}{2\pi}\int_{-\pi}^{\pi}\!\frac{dq_{j}}{2\pi}\right]\e^{i\omega\tau(\cos(q_{1})+\cdots+\cos(q_{N}))}\nonumber\\
    &\quad
      \times\langle0|\widetilde{a}_{p_{N}}\cdots\widetilde{a}_{p_{1}}\widetilde{a}_{q_{1}}^{\dagger}\cdots\widetilde{a}_{q_{N}}^{\dagger}|0\rangle\e^{ip_{1}x_{1}+\cdots+ip_{N}x_{N}-iq_{1}y_{1}-\cdots-iq_{N}y_{N}}\nonumber\\
    &=\sum_{\sigma\in S_{N}}(\pm1)^{\#\sigma}\prod_{j=1}^{N}\e^{i\frac{\pi}{2}|x_{j}-y_{\sigma(j)}|}J_{|x_{j}-y_{\sigma(j)}|}(\omega\tau).\label{eq:A.15}
  \end{align}
  This equation can be used to construct the time-evolution kernels
  for free identical walkers on a circle, the half line, and a finite
  interval.
\end{itemize}

\subsection{Tight-binding model on a circle}
\label{appendix:A.2}
Let us next consider the tight-binding model for free spinless
particles on the periodic lattice $\{1,2,\cdots,L\pmod L\}$ subject to
the twisted boundary condition $a_{x+L}=\e^{i\theta}a_{x}$. As we
shall see shortly, the following Hamiltonian operator yields the
desired results:
\begin{align}
  H=-\frac{\omega}{2}\sum_{x=1}^{L}\left(a_{x+1}^{\dagger}a_{x}+a_{x}^{\dagger}a_{x+1}\right),\quad\text{where}\quad a_{L+1}\equiv\e^{i\theta}a_{1}.\label{eq:A.16}
\end{align}
In the following, we assume that $\theta$ ranges from $0$ to $2\pi$.

In order to compute the time-evolution kernel, we first have to
diagonalize the Hamiltonian operator \eqref{eq:A.16}, which can be
done by using the mode expansion. Under the twisted boundary
condition, the annihilation operator can be expanded into the
following:
\begin{align}
  a_{x}=\frac{1}{\sqrt{L}}\sum_{p=0}^{L-1}\widetilde{a}_{p}\e^{i\frac{2p\pi+\theta}{L}x},\label{eq:A.17}
\end{align}
where $\widetilde{a}_{p}$ and $\widetilde{a}_{p}^{\dagger}$ satisfy
$[\widetilde{a}_{p},\widetilde{a}_{q}^{\dagger}]_{\mp}=\delta_{p,q}$
and $[\widetilde{a}_{p},\widetilde{a}_{q}]_{\mp}=0$ for any
$p,q\in\{0,1,\cdots,L-1\}$. By substituting eq.~\eqref{eq:A.17} into
eq.~\eqref{eq:A.16}, we find that the Hamiltonian operator is
diagonalized as follows:
\begin{align}
  H=\sum_{p=0}^{L-1}\varepsilon_{p}\widetilde{a}_{p}^{\dagger}\widetilde{a}_{p},\label{eq:A.18}
\end{align}
where $\varepsilon_{p}=-\omega\cos(\frac{2p\pi+\theta}{L})$ is the
single-particle energy eigenvalue on the periodic lattice.

Now it is easy to compute the time-evolution kernel in the
one-particle sector. A straightforward calculation gives
\begin{align}
  U_{\tau}^{[\theta]}(x,y)
  &=\langle x|\e^{-iH\tau}|y\rangle\nonumber\\
  &=\langle0|a_{x}\e^{-iH\tau}a_{y}^{\dagger}|0\rangle\nonumber\\
  &=\frac{1}{L}\sum_{p=0}^{L-1}\sum_{q=0}^{L-1}\langle0|\widetilde{a}_{p}\e^{-iH\tau}\widetilde{a}_{q}^{\dagger}|0\rangle\e^{i\frac{2p\pi+\theta}{L}x-i\frac{2q\pi+\theta}{L}y}\nonumber\\
  &=\frac{1}{L}\sum_{p=0}^{L-1}\sum_{q=0}^{L-1}\e^{-i\varepsilon_{q}\tau}\langle0|\widetilde{a}_{p}\widetilde{a}_{q}^{\dagger}|0\rangle\e^{i\frac{2p\pi+\theta}{L}x-i\frac{2q\pi+\theta}{L}y}\nonumber\\
  &=\frac{1}{L}\sum_{p=0}^{L-1}\e^{i\omega\tau\cos(\frac{2p\pi+\theta}{L})}\e^{i\frac{2p\pi+\theta}{L}(x-y)},\label{eq:A.19}
\end{align}
where we have used
$\e^{-iH\tau}\widetilde{a}_{q}^{\dagger}|0\rangle=\e^{-i\varepsilon_{q}\tau}\widetilde{a}_{q}^{\dagger}|0\rangle$
in the fourth line and
$\langle0|\widetilde{a}_{p}\widetilde{a}_{q}^{\dagger}|0\rangle=\delta_{p,q}$
in the last line. Notice that eq.~\eqref{eq:A.19} is the summation
over the energy spectrum. In order to obtain the summation over
winding numbers, we therefore have to perform a resummation, which can
be done by using eq.~\eqref{eq:A.7}. By substituting
$\e^{i\omega
  t\cos(\frac{2p\pi+\theta}{L})}=\sum_{m\in\mathbb{Z}}\e^{i\frac{\pi}{2}|m|}J_{|m|}(\omega
t)\e^{-im\frac{2p\pi+\theta}{L}}$ into eq.~\eqref{eq:A.19} and using
the orthogonal relation
$(1/L)\sum_{p=0}^{L-1}\e^{i\frac{2p\pi+\theta}{L}(x-y-m)}=\e^{in\theta}\delta_{m,x-y-nL}$
($n\in\mathbb{Z}$), we find that the time-evolution kernel
\eqref{eq:A.19} can be put into the following alternative equivalent
form:\footnote{The case $\theta=0$ was noted in
  ref.~\cite{Ahmadi:2002}.}
\begin{align}
  U_{\tau}^{[\theta]}(x,y)=\sum_{n=-\infty}^{\infty}\e^{in\theta}\e^{i\frac{\pi}{2}|x-y-nL|}J_{|x-y-nL|}(\omega\tau),\label{eq:A.20}
\end{align}
which exactly coincides with eq.~\eqref{eq:15} with
$\widetilde{U}_{\tau}(\cdot,\cdot)$ given by eq.~\eqref{eq:A.8}. This
sample computation implies that there is an equivalence (or duality)
between the summation over energy spectrum and the summation over
particle's trajectories, which is the heart of the trace formula in
harmonic analysis and representation theory (see, e.g.,
ref.~\cite{Marklof:2008}). In this respect, one could say that our
formula is a version of the trace formula in lattice geometry.

Although we omit the details, it is not difficult to show that the
resolvent kernel, the canonical density matrix, and the time-evolution
kernel for $N$ identical particles all coincide with the universal
formulas.

We note in closing that the parameter $\theta$ can be removed from the
twisted boundary condition under the gauge transformation
$a_{x}\mapsto
V_{\theta}a_{x}V_{\theta}^{-1}=\e^{i\frac{\theta}{L}x}a_{x}$, where
$V_{\theta}$ is a unitary operator given by
$V_{\theta}=\exp(-i\frac{\theta}{L}\sum_{x=1}^{L}xa_{x}^{\dagger}a_{x})$
(see, e.g., ref.~\cite{Lin:2022zbh}). In fact, a straightforward
calculation gives
\begin{align}
  V_{\theta}HV_{\theta}^{-1}=-\frac{\omega}{2}\sum_{x=1}^{L}\left(\e^{-i\theta/L}a_{x+1}^{\dagger}a_{x}+\e^{+i\theta/L}a_{x}^{\dagger}a_{x+1}\right),\quad\text{where}\quad a_{L+1}\equiv a_{1}.\label{eq:A.21}
\end{align}
The time-evolution kernel in the one-particle sector for this
Hamiltonian coincides with eq.~\eqref{eq:A.20} up to a phase factor
$\e^{i\frac{\theta}{L}(x-y)}$ and hence is physically equivalent.

\subsection{Tight-binding model on the half line}
\label{appendix:A.3}
Let us next consider the tight-binding model on the semi-infinite
lattice $\{1,2,\cdots\}$ with the boundary condition
$a_{0}=e^{i\phi}a_{1}$, where $\phi\in\{0,\pi\}$. The Hamiltonian
operator that ensures this boundary condition is given by
\begin{align}
  H=-\frac{\omega}{2}\sum_{x=1}^{\infty}\left(a_{x+1}^{\dagger}a_{x}+a_{x}^{\dagger}a_{x+1}\right)-\frac{\omega}{2}\e^{i\phi}a_{1}^{\dagger}a_{1}.\label{eq:A.22}
\end{align}
By substituting the mode expansion
\begin{align}
  a_{x}=\int_{0}^{\pi}\!\frac{dp}{2\pi}\,\widetilde{a}_{p}\left(\e^{-ipx}+\e^{i\phi}\e^{-ip(1-x)}\right),\label{eq:A.23}
\end{align}
we get the following diagonalized Hamiltonian operator:
\begin{align}
  H=\int_{0}^{\pi}\!\frac{dp}{2\pi}\,\varepsilon_{p}\widetilde{a}_{p}^{\dagger}\widetilde{a}_{p},\label{eq:A.24}
\end{align}
where $\varepsilon_{p}=-\omega\cos(p)$ is the single-particle energy
eigenvalue. The time-evolution kernel for a single walker is given by
\begin{align}
  U_{\tau}^{[\phi]}(x,y)
  &=\langle0|a_{x}\e^{-iH\tau}a_{y}^{\dagger}|0\rangle\nonumber\\
  &=\int_{0}^{\pi}\!\frac{dp}{2\pi}\e^{-i\varepsilon_{p}\tau}\left(\e^{-ipx}+\e^{i\phi}\e^{-ip(1-x)}\right)\left(\e^{ipy}+\e^{i\phi}\e^{ip(1-y)}\right)\nonumber\\
  &=\int_{-\pi}^{\pi}\!\frac{dp}{2\pi}\e^{i\omega\tau\cos(p)}\left(\e^{ip(x-y)}+\e^{i\phi}\e^{ip(x-1+y)}\right)\nonumber\\
  &=\e^{i\frac{\pi}{2}|x-y|}J_{|x-y|}(\omega\tau)+\e^{i\phi}\e^{i\frac{\pi}{2}|x-1+y|}J_{|x-1+y|}(\omega\tau),\label{eq:A.25}
\end{align}
which exactly coincides with eq.~\eqref{eq:18}. Other quantities can
be calculated in a similar way and coincide with the universal
formulas.

We note that the model that satisfies the Dirichlet boundary condition
$a_{x}=0$ at $x=0$ is described by the Hamiltonian operator
$H=-(\omega/2)\sum_{x=1}^{\infty}(a_{x+1}^{\dagger}a_{x}+a_{x}^{\dagger}a_{x+1})$. In
this case, the time-evolution kernel coincides with another formula
discussed in example 2 in section~\ref{section:3.1}.

\subsection{Tight-binding model on a finite interval}
\label{appendix:A.4}
Let us finally quickly study the tight-binding model on the finite
lattice $\{1,2,\cdots,L\}$ with the boundary conditions
$a_{0}=\e^{i\phi}a_{1}$ and $a_{L+1}=\e^{i(\theta+\phi)}a_{L}$, where
$\theta,\phi\in\{0,\pi\}$. The Hamiltonian operator is given by
\begin{align}
  H=-\frac{\omega}{2}\sum_{x=1}^{L-1}\left(a_{x+1}^{\dagger}a_{x}+a_{x}^{\dagger}a_{x+1}\right)-\frac{\omega}{2}\e^{i\phi}a_{1}^{\dagger}a_{1}-\frac{\omega}{2}\e^{i(\theta+\phi)}a_{L}^{\dagger}a_{L}.\label{eq:A.26}
\end{align}
This operator can be diagonalized by using the following mode
expansions:
\begin{align}
  a_{x}=
  \begin{dcases}
    \frac{1}{\sqrt{L}}\widetilde{a}_{0}+\frac{1}{\sqrt{2L}}\sum_{p=1}^{L-1}\widetilde{a}_{p}\left(\e^{-i\frac{2p\pi}{2L}x}+\e^{-i\frac{2p\pi}{2L}(1-x)}\right)&\text{for}\quad\theta=0~~\&~~\phi=0;\\
    \frac{1}{\sqrt{2L}}\sum_{p=1}^{L-1}\widetilde{a}_{p}\left(\e^{-i\frac{2p\pi}{2L}x}-\e^{-i\frac{2p\pi}{2L}(1-x)}\right)+\frac{1}{\sqrt{L}}\widetilde{a}_{L}(-1)^{x}&\text{for}\quad\theta=0~~\&~~\phi=\pi;\\
    \frac{1}{\sqrt{2L}}\sum_{p=0}^{L-1}\widetilde{a}_{p}\left(\e^{-i\frac{2p\pi+\theta}{2L}x}+\e^{i\phi}\e^{-i\frac{2p\pi+\theta}{2L}(1-x)}\right)&\text{otherwise}.\\
  \end{dcases}\label{eq:A.27}
\end{align}
In fact, by substituting these into eq.~\eqref{eq:A.26} we find
\begin{align}
  H=
  \begin{dcases}
    \sum_{p=1}^{L}\varepsilon_{p}\widetilde{a}_{p}^{\dagger}\widetilde{a}_{p}&\text{for}\quad\theta=0~~\&~~\phi=\pi;\\
    \sum_{p=0}^{L-1}\varepsilon_{p}\widetilde{a}_{p}^{\dagger}\widetilde{a}_{p}&\text{otherwise},\\
  \end{dcases}\label{eq:A.28}
\end{align}
where $\varepsilon_{p}=-\omega\cos(\frac{2p\pi+\theta}{2L})$ for any
$\theta,\phi\in\{0,\pi\}$. It is not difficult to show that the
time-evolution kernel for a single walker can be put into the
following expression irrespective of the values of $\theta$ and
$\phi$:
\begin{align}
  U_{\tau}^{[\theta,\phi]}(x,y)
  &=\langle0|a_{x}\e^{-iH\tau}a_{y}^{\dagger}|0\rangle\nonumber\\
  &=\frac{1}{2L}\sum_{p=0}^{2L-1}\e^{i\omega\tau\cos(\frac{2p\pi+\theta}{2L})}\left(\e^{i\frac{2p\pi+\theta}{2L}(x-y)}+\e^{i\phi}\e^{i\frac{2p\pi+\theta}{2L}(x-1+y)}\right).\label{eq:A.29}
\end{align}
Note that this is the summation over the energy spectrum. However, as
was done in appendix \ref{appendix:A.2}, this summation can be
rewritten into the following summation over the bouncing numbers off
the boundaries:
\begin{align}
  U_{\tau}^{[\theta,\phi]}(x,y)=\sum_{n=-\infty}^{\infty}\left[\e^{in\theta}\e^{i\frac{\pi}{2}|x-y-2nL|}J_{|x-y-2nL|}(\omega\tau)+\e^{in\theta}\e^{i\phi}\e^{i\frac{\pi}{2}|x-1+y-2nL|}J_{|x-1+y-2nL|}(\omega\tau)\right],\label{eq:A.30}
\end{align}
which exactly coincides with the universal formula \eqref{eq:21}.

If one wants to study the model that satisfies the Dirichlet boundary
conditions $a_{x}=0$ at $x=0$ and $x=L+1$, one should use
$H=-(\omega/2)\sum_{x=1}^{L-1}(a_{x+1}^{\dagger}a_{x}+a_{x}^{\dagger}a_{x+1})$. In
this case, the time-evolution kernel coincides with another formula
discussed in example 3 in section~\ref{section:3.1}.

\printbibliography[heading=bibintoc]
\end{document}